\documentclass[11pt]{article}
\usepackage{epsfig}
\usepackage{amssymb}

\newcommand{\etal}{{\em et al.\/}}
\newcommand{\tH}{{\rm th}}

\newcommand{\ie}{{\em i.e.\/}}

\newcommand{\viz} {{\em viz\/}}

\newcommand{\calA}{{\cal A}}

\newcommand{\calD}{{\cal D}}

\newcommand{\calH}{{\cal H}}

\newcommand{\calL}{{\cal L}}
\newcommand{\calM}{{\cal M}}

\newcommand{\maly}[1]{{\scriptscriptstyle#1}}
\newcommand{\malyzero}{{\maly{0}}}
\newcommand{\malyone}{{\maly{1}}}
\newcommand{\malytwo}{{\maly{2}}}

\newcommand{\ssb}[1]{{_{#1}}}
\newcommand{\ssp}[1]{{^{#1}}}

\newcommand{\parend}[1]{{\left( #1  \right) }}

\newcommand{\braced}[1]{{\left\{ #1  \right\} }}

\newcommand{\floor}[1]{{\left\lfloor #1\right\rfloor}}

\newcommand{\barred}[1]{{\left|#1\right|}}

\newcommand{\spaced}[1]{{\ #1\ }}
\newcommand{\romn}[1]{{\mbox{\rm #1}}}
\newcommand{\spacedromn}[1]{\spaced{\romn{#1}}}
\newcommand{\italicized}[1]{{\mbox{\it #1}}}

\newcommand{\leftbracedtwo}[2]{\left\{\begin{array}{l}
                                        #1 \\ #2
                                    \end{array}\right.
                           }

\newcommand{\ignore}[1]{}

\newtheorem{theorem}{Theorem}

\newtheorem{observation}{Observation}

\newtheorem{lemma}{Lemma}

\newenvironment{proof}{{\noindent \it Proof:\/}}{\hfill $\square$}

\newcommand{\half}{{\mbox{$\frac{1}{2}$}}}

\newcommand{\opt}{{\mbox{\scriptsize \sc opt}}}

\newcommand{\adv}{{\mbox{\it adv}}}
\newcommand{\alg}{{\mbox{\it alg}}}
\newcommand{\cost}{{\mbox{\it cost\/}}}

\newcommand{\reals}{{\bf R}}

\usepackage[dvips,bookmarks=false]{hyperref}

\title{\bf T-Theory Applications to Online Algorithms \\for the Server Problem}

\author
        {
        Lawrence L. Larmore
        \thanks{School of Computer Science,
        University of Nevada,
        Las Vegas, NV 89154.
         Email: {\tt larmore@cs.unlv.edu}.
 Research supported by NSF grant CCR-0312093.
           }
        \and
        James A. Oravec
        \thanks{School of Computer Science,
        University of Nevada,
        Las Vegas, NV 89154.
        Email: {\tt james.oravec@gmail.com}.
 Research supported by NSF grant CCR-0312093.
                }
        }

\begin{document}

\maketitle

\begin{abstract}
{\bf {\large {\em
Although largely unnoticed by the online algorithms community,
T-theory, a field of discrete mathematics, has contributed to the
development of several online algorithms for the $k$-server problem.
A brief summary of the $k$-server problem, and some important
application concepts of T-theory, are given.
Additionally, a number of known $k$-server results are restated using
the established terminology of T-theory. Lastly, a previously
unpublished 3-competitiveness proof, using T-theory,
for the {\sc Harmonic} algorithm for two servers is presented.
} } }
\end{abstract} 

{\bf Keywords/Phrases\/}:
   { Double Coverage},
   { Equipoise},
   \mbox{Harmonic},
   Injective Hull,
   Isolation Index,
   Prime Metric,
   { Random Slack},
   T-Theory,
   Tight Span,
   Trackless Online Algorithm
\eject
{
\small
\tableofcontents
}
\eject

\section{\bf Introduction}\label{sec: new introduction}

The $k$-{\em server problem\/}
was introduced by Manasse, McGeoch, and Sleator \cite{MaMcSl88},
while {\em T-theory\/} was introduced by John Isbell \cite{Isbell64}
and independently rediscovered by Andreas Dress \cite{Dress84,Dress89}.
The communities of
researchers in these two areas have had little interaction.
The {\em tight span\/},
a fundamental construction of T-theory, was later defined independently, using
different notation, by Chrobak and Larmore,
who were unaware of the work of Isbell, Dress, and others.

Bartal \cite{Bartal94b}, 
Chrobak and Larmore \cite{ChrLar91B,ChrLar91C,ChrLar91A,ChrLar92C},
and Teia \cite{Teia93b}, have used the tight span concept to
obtain results for the
$k$-{\it \em server problem\/}.
In this paper, we summarize those results,
using the standard notation of
{\it \em T-theory\/}.
We then suggest ways to use {\it \em T-theory\/}
to obtain additional results for the $k$-{\it \em server problem\/}.

\subsection{\bf The $k$-Server Problem}\label{subsec: new k server}

Let $M$ be a metric space, in which there are $k$ identical mobile
{\em servers\/}.
At each time step a request point $r \in M$ is given,
and one server must move to $r$ to {\em serve\/} the request.
The measure of cost is the total distance
traveled by the servers over the entire sequence of requests.

An {\em online algorithm\/} is an algorithm which must decide on some
outputs before knowing all inputs.
  Specifically, an {\it\em online algorithm\/}
for the server problem must decide
which server to move to a given request point, without knowing
the sequence of future requests, as opposed to an {\em offline\/}
algorithm, which knows all requests in advance.

For any constant $C \ge 1$
 we say that an {\it\em online algorithm\/} $\calA$
 for the server problem\footnote{Or for any of a large number of other
 {\em online\/} problems.} is {\em $C$-competitive\/}
 if there exists a constant $K$ such that,
 for any request sequence $\varrho$,
 where $\cost_\ssb{\opt}(\varrho)$
 is the optimum cost for serving that sequence:
$$\cost_\ssb{\calA}(\varrho) \le C\cdot\cost_\ssb{\opt}(\varrho) + K$$
If $\calA$ is randomized, the expected cost $E\cost_\ssb{\calA}(\varrho)$
is used instead of $\cost_\ssb{\calA}(\varrho)$.

The {\em $k$-server conjecture\/}, posed by Manasse, McGeoch, and Sleator
\cite{MaMcSl88}, is that there
is a deterministic $k$-competitive online algorithm for the $k$-server
problem in an arbitrary metric space.
Since its introduction by Manasse \etal, substantial
work has been done on the $k$-{\it \em server problem}
\cite{ AlKaPW95, BalShe93, Bartal94b, BaChLa98, BaChLa00, BaChNR02,
BarGro00, BarMen04, BarRos92, BarKou04, BeChLa99, BeChLa02, BeKaTa90,
BlKaRS92, BlKaRS00, BorElY98, ChrLar92B, ChrLar98, ChLaLR97, ChrSga04,
ElYKle95, EpImSt03, FiRaRa94, FiRaRS94, FiaRic94, Grove91, IraRub91,
KaRaRa94, KouPap95A, Koutso99, Teia93b, Young94}.
The $k$-server conjecture remains open, except for special cases, and for
$k=2$ for all cases.

It is traditional to analyze the competitiveness of an online algorithm
by imagining the existence of an {\em adversary\/}, who creates the
request sequence, and must also serve that same sequence.  Since we
assume that the adversary has unlimited computational power, it will
serve the request sequence optimally; thus, competitiveness can be
calculated by comparing the cost incurred by the online algorithm to
the cost incurred by that adversary.
We refer the reader to Chapter 4 of \cite{BorElY98} for an extensive
discussion of adversarial models.

Throughout this paper, we will let $s_\ssb{1}, \ldots, s_\ssb{k}$ denote the
algorithm's
servers, and also, by an abuse of notation, the points where the servers
are located.  Similarly, we will let $a_\ssb{1}, \ldots, a_\ssb{k}$ denote
both the adversary's servers and the points where they are located.
We will also let $r$ be the request point.

\subsection{\bf Memoryless, Fast, and Trackless $k$-Server Algorithms}
\label{subsec: memoryless fast trackless}

Let $\calA$ be an online algorithm for the $k$-server problem.
$\calA$ is called {\em memoryless\/}
if its only memory between steps
is the locations of its own servers.  When a new request is received,
$\calA$ makes a decision, moves its servers, and then forgets all information
except the new locations of the servers.

$\calA$ is called {\em fast\/} if, after each request,
$\calA$ can make
its decision using $O(1)$ operations,
where computing the distance between two points counts as one operation.

$\calA$ is called {\em trackless\/}
if $\calA$ initially knows only the distances between its various servers.
When $\calA$ receives a request,
it is only told the distances between that request and each of its servers.
$\calA$'s only allowed output is an instruction to move a specific server to
the request point.
 $\calA$ may not have any naming system for points.
Thus, it cannot tell how close a given request is to any point
on which it does not currently have a server. 
See \cite{BeiLar00} for further discussion of tracklessness.

\subsection{\bf The Lazy Adversary}\label{subsec: new lazy adversary}

The {\em lazy adversary\/} is an adversary that always makes a request that
costs it nothing to serve, but which forces the algorithm to pay, if
such a request is possible.  For the $k$-server problem, the lazy
adversary always requests a point
where one of its servers is located, provided the algorithm has no server
at that point.  When all algorithm servers are at the same points
as the adversary servers, the lazy adversary may move one of its servers
to a new point.
Thus, the lazy adversary never has more than one server that is in a
position different from that of an algorithm server.
Some online algorithms, such as {\sc handicap},
introduced in Section \ref{sec: new handicap}, perform better against
the lazy adversary than against an adversary without that restriction.

\subsection{\bf T-Theory and its Application to the $k$-Server
Problem}\label{subsec: new ttheory}

Since the pioneering work by Isbell and Dress, there have
been many contributions to the field of {\it\em T-theory\/}
\cite{BanChe96, BanChe98, BanDre92A, BuBuIv01, Chepoi97, Christo97, DrHuKM01,
DrHuMo01, DrHuMo02, DrHuMo96, DrMoTe96, DreSch87, Herrma05, HuKoMo04, HuKoMo05,
HuKoMo06, Isbell64, KoMoTo98, KoMoTo00, SVTSDK96, StuYu04}.
The original motivation for the development of {\it\em T-theory\/},
and one of its most important application areas, is
{\em phylogenetic analysis}, the problem of constructing a
phylogenetic tree showing relationships among species or languages
\cite{BarGue91,SemSte03}.

It was first discovered by Chrobak and Larmore \cite{ChrLar91B}
that T-theory can aid in the competitive analysis of online algorithms
for the $k$-server problem.  Since then, work by Teia \cite{Teia93b}
and Bartal \cite{Bartal94b}, and additional work by Chrobak and Larmore
 \cite{ChrLar91C,ChrLar92C} have made use of T-theory concepts to
obtain $k$-server results.

Many proofs of results in the area of the $k$-server problem require
lengthy case-by-case analysis.  T-theory can help guide this process
by providing a natural way to break a proof or a definition into cases.
This can be seen in this paper in the definitions of {\sc balance slack}
\S \ref{subsec: new balance slack}
and {\sc handicap}
\S \ref{subsec: new handicap}, and in the proof of 3-competitiveness of
{\sc harmonic} for $k=2$, in Section \ref{sec: new harmonic}.

In a somewhat different use of T-theory,
the {\em tight span\/} algorithm and {\sc equipoise} make use of the
{\em virtual server method\/}
discussed in Section \ref{sec: new virtual}.  These algorithms
move servers virtually in the tight span of a metric space.

\subsection{\bf Overview of the Paper}\label{subsec: new overview}

In Section \ref{sec: new ttheory},
we give some elementary constructions from T-theory that are used
in applications to the server problem.  We provide illustrations and
pseudo code for a number of algorithms that we describe.

In Section \ref{sec: new virtual}, we give the virtual server
construction, which is used for the {\it\em tight span\/} algorithm as
well as for {\sc equipoise}.
In \S \ref{subsec: new tree}, we describe the {\em tree algorithm\/}
({\sc tree}) of \cite{ChrLar91A}, which forms the basis of a number
of the other server algorithms described in this paper.
In \S \ref{subsec: new bartal}, we describe the Bartal's
 {\em Slack Coverage\/}
algorithm for 2 servers in a Euclidean space \cite{Bartal94b}
 in terms of T-theory.

In Section \ref{sec: new balance},
we discuss {\em balance\/} algorithms for the $k$-server problem in
terms of $T$-theory.
In Section \ref{sec: new server tight span} we describe
the {\it\em tight span algorithm} \cite{ChrLar91B}
 and {\sc equipoise} \cite{ChrLar92C} in terms of T-theory.
In \mbox{Section \ref{sec: new handicap}}, we present a description of
Teia's algorithm {\sc handicap}.
In \S \ref{subsec: new random slack}
we describe how the algorithm
{\sc random slack} is defined using T-theory.
In \S \ref{subsec: new harmonic analysis}, we present a T-theory based
proof that {\sc harmonic}  \cite{RagSni89}
 is 3-competitive for $k=2$.
In Section \ref{sec: new summary future}, we discuss possible future
uses of T-theory for the $k$-server problem.

We present a simplified proof that {\sc handicap} is $k$-competitive
against certain adversaries (Theorem \ref{thm: handicap}),
based on the proof in Teia's disseration \cite{Teia93b}.
We also give a previously unpublished proof that
{\sc harmonic} is 3-competitive
for $k=2$.

\section{\bf Elementary T-Theory Concepts}\label{sec: new ttheory}

In keeping with the usual practice of
T-theory papers, we extend the meaning of the term
{\em metric\/} to incorporate what is commonly called a {\em pseudo-metric\/}.
That is, we define a {\em metric\/} on a set $X$
to be a function $d:X\times X\to\reals$ such that
\begin{enumerate}
\item $d(x,x) = 0$ for all $x\in X$
\item $d(x,y) = d(y,x)$ for all $x,y\in X$
\item $d(x,y) + d(y,z) \ge d(x,z)$ for all $x,y,z\in X$
{\bf(Triangle Inequality)}
\end{enumerate}
We say that $d$ is a {\em proper metric\/} if, in addition,
$d(x,y) > 0$ whenever $x\ne y$.
We also adopt the usual practice of abbreviating a metric space $(X,d)$
as simply $X$, if $d$ is understood.

\subsection{\bf Injective Spaces and the
Tight Span}\label{subsec: new injective}

Isbell \cite{Isbell64} defines a metric space $M$ to be {\em injective\/}
if, for any metric space $Y\supseteq M$, there is a non-expansive retraction
of $Y$ onto $M$, \ie, a map
$r:Y\to M$ which is the identity on $M$, where $d(r(x),r(y))\le d(x,y)$
for all $x,y\in Y$.
The real line, the {\em Manhattan plane\/},
\ie, the plane $\reals^2$ with the $\calL_1$ (sum of norms) metric,
and $\reals^n$ with the $\calL_\infty$ (sup-norm) metric, where the
distance between
$\parend{x_\ssb{1},\ldots ,x_\ssb{n}}$ and
$\parend{y_\ssb{1},\ldots ,y_\ssb{n}}$ is
$\max_{1\le i\le n}\barred{x_\ssb{i}-y_\ssb{i}}$,
are injective.
No Euclidean space of dimension more than one is injective.

\begin{figure}[ht!]
\centerline{\epsfig{file=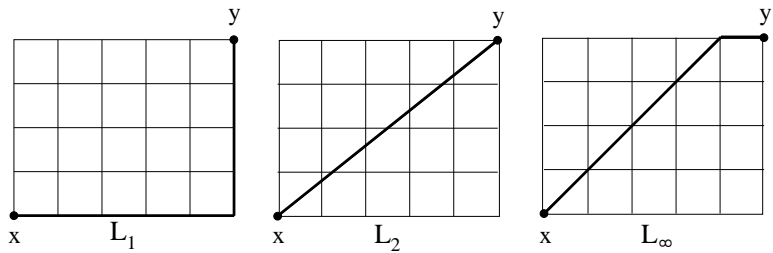, width= 4in}}
\caption{\small \bf Illustration of $L_1$, $L_2$, and $L_{\infty}$
metrics on the plane.
The distance between $x$ and $y$ is 9 with the $L_1$ metric,
$\sqrt{41}$ with the $L_2$ metric and 5 with the $L_{\infty}$ metric.
}
\label{fig: L1 L2 and Linf}
\end{figure}

The {\em tight span\/} $T(X)$ of a metric space $X$, which we
formally define below,
is characterized by a {\em universal property:\/}
up to isomorphism, $T(X)$ is the unique minimal injective metric space
that contains $X$.  Thus, $X=T(X)$ if and only if $X$ is injective.

Isbell \cite{Isbell64} was the first to construct $T(X)$,
which he called the {\em injective hull\/} of $X$.
Dress \cite{Dress84} independently developed the same construction,
naming it the {\it\em tight span} of $X$.
Still later, Chrobak and Larmore also independently developed the
tight span, which they called the {\em abstract convex hull\/} of $X$.

We now give a formal construction of $T(X)$.  Let
\begin{eqnarray}
P(X) = \braced{f\in\reals^X\ |\ f(x)+f(y)\ge d(x,y)\spacedromn{for all}
x,y\in X}
\label{eqn: new p of x}
\end{eqnarray}
where $\reals^X$ is the set of all functions $f:X\to \reals$,
and let $T(X)\subseteq P(X)$ be the set of those functions which
are minimal with respect to pointwise partial order.
$T(X)$ is a metric space where distance is given by 
the sup-norm metric, \ie,
If $f,g\in T(X)$, we define $d(f,g)=\sup_{x\in X}\barred{f(x)-g(x)}$.
If $X$ is finite, then $P(X)$ is also a metric space under the
sup-norm metric, and is called 
the {\em associated polytope\/} of $X$ \cite{KoMoTo98}.
There is a canonical embedding\footnote{In this paper, {\em embedding\/}
will mean {\em isometric embedding}.} of $X$ into $T(X)$.
For any $x\in X$, let $h_\ssb{x}\in T(X)$ be the
function where $h_\ssb{x}(y)=d(x,y)$
for all $y$.  By an abuse of notation, we identify each $x$ with $h_\ssb{x}$,
and thus say $X\subseteq T(X)$.

If $X$ has cardinality $n$,
then $P(X) \subseteq \reals^X\cong\reals^n$.
For any $x,y\in X$, let $D_{x,y}\subseteq \reals^X$ be the 
half-space defined by the inequality $f(x)+f(y) \ge d(x,y)$,
and let $H_{x,y}\subseteq \reals^X$ be the boundary of $D_{x,y}$,
the hyperplane defined by the equation $f(x)+f(y) = d(x,y)$, which
we call a {\em bounding hyperplane\/} of $P(X)$.
Then $P(X) = \bigcap_{x,y\in X}D_{x,y}$ is
an unbounded convex polytope of \mbox{dimension $n$},
 and $T(X)$ is the union of all the bounded faces of $P(X)$.

The definition of {\em convex subset\/} of a metric space is not
consistent with the definition of {\em convex subset\/} of a vector space
over the real numbers.
$T(X)$ is a convex subset
of $P(X)$, if $P(X)$ is considered to be a metric space; but
$T(X)$ is not generally a convex subset of $\reals^X$ if $\reals^X$
is considered to be a vector space over $\reals$.

\begin{figure}[ht!]
\centerline{\epsfig{file=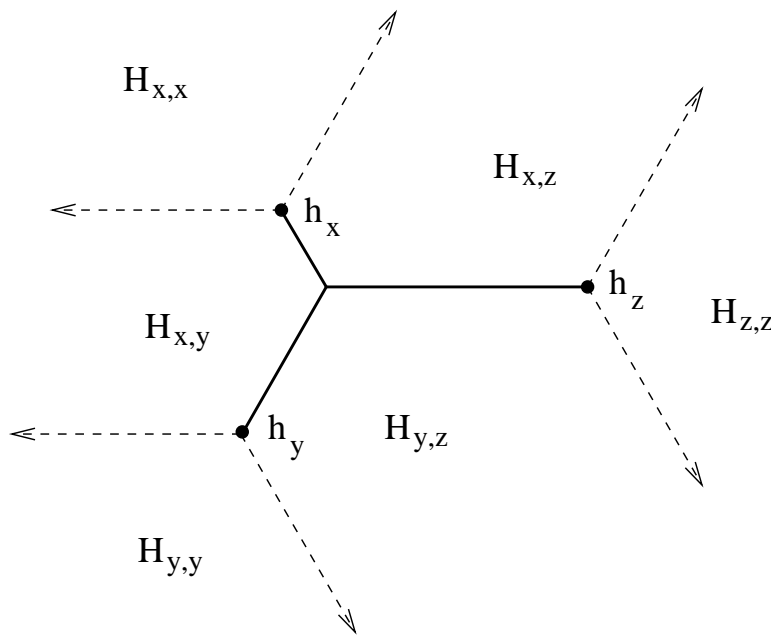, width= 4in}}
\caption{\small \bf A two-dimensional projection of the
three-dimensional complex $P(X)$,
in the case where $X$ is the 3-4-5 triangle.  $T(X)$ is the subcomplex
consisting of the vertices and the bold line segments.
}
\label{fig: example tight span A}
\end{figure}

\begin{figure}[ht!]
\unitlength1cm
\begin{minipage}[t]{5.5cm}
\centerline{\epsfig{file=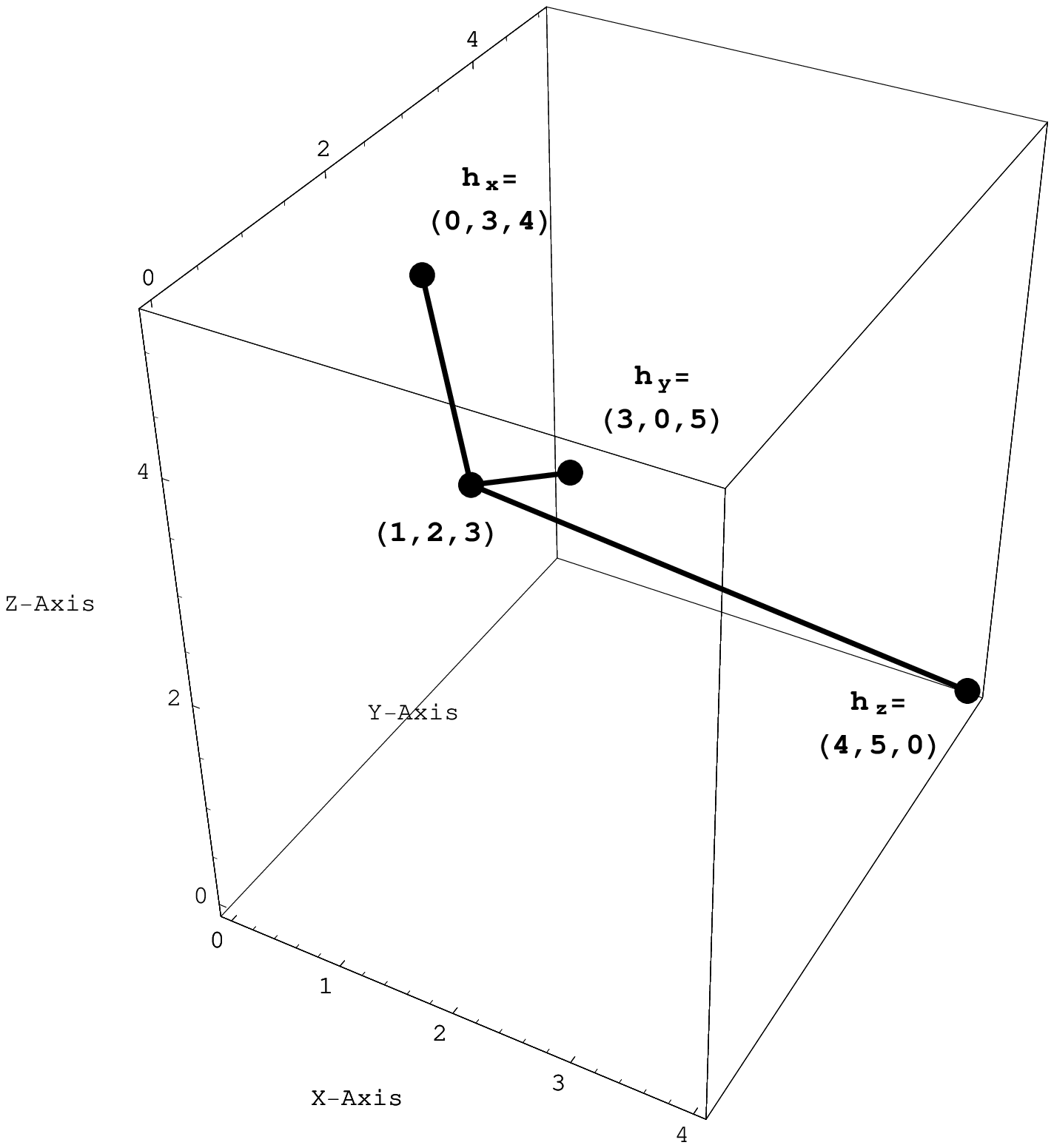, width= 2.5in}}
\caption{\small \bf  A view of the tight span of
the 3-4-5 triangle, embedded in $(\reals^3,\calL_\infty)$.
}
\label{fig: example tight span B}
\end{minipage}
\hfill
\begin{minipage}[t]{6cm}
\centerline{\epsfig{file=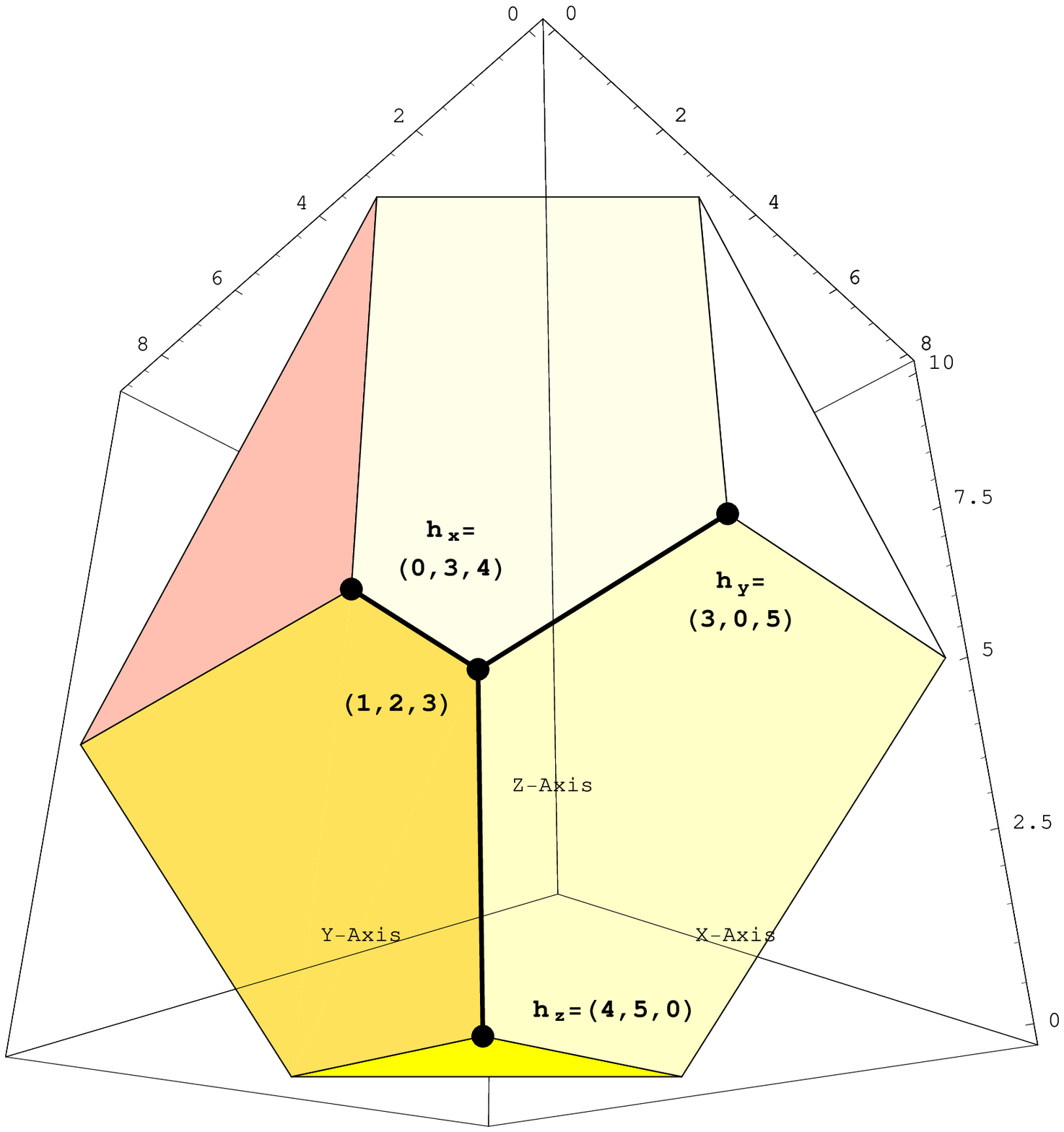, width= 3.1in}}
\caption{\small \bf A view of the associated polytope of the 3-4-5
triangle, with the tight span in bold.
}
\label{fig: example tight span C}
\end{minipage}
\end{figure}

Dress proves \cite{Dress89} that
the tight span of a metric space of cardinality $n$
is a {\em cell complex\/}, where each
cell is a polytope of dimension at most $\floor{\frac{n}{2}}$.
In Figures \ref{fig: example tight span A},
\ref{fig: example tight span B},
and \ref{fig: example tight span C},
we give an example of the tight span of the 3-4-5 triangle.
Let $X=\braced{x,y,z}$, where
$d(x,y)=3$, $d(x,z)=4$, and $d(y,z)=5$.  The vertices of $T(X)$,
represented as 3-tuples in $\reals^3\cong\reals^X$, are

\begin{eqnarray*}
 h_x = (0,3,4) &=& H_{x,x}\cap H_{x,y}\cap H_{x,z}\\
 h_y = (3,0,5) &=& H_{x,y}\cap H_{y,y}\cap H_{y,z}\\
 h_z = (4,5,0) &=& H_{x,z}\cap H_{y,z}\cap H_{z,z}\\
 (1,2,3) &=& H_{x,y}\cap H_{x,z}\cap H_{y,z}
\end{eqnarray*}

Figure
\ref{fig: example tight span A} shows a projection of $P(X)$ in two dimensions.
The boundary of $P(X)$ consists of four vertices, three bounded edges,
six unbounded edges, and six unbounded 2-faces.
$T(X)$ is the union of the bounded edges.
Figure \ref{fig: example tight span B} is a perspective showing $T(X)$
in $\reals^3$, which we endow with the $\calL_\infty$ metric.
Figure \ref{fig: example tight span C} is a rendering of the polytope obtained
by intersecting $P(X)$ with a half space.

\subsection{\bf The Isolation Index
 and the Split Decomposition}\label{subsec: new isolation split}
Let $(X,d)$ be a metric space.  
If $A,B\subseteq X$ are non-empty subsets of $X$,
Bandelt and Dress \cite{BanDre92A} (page 54) define
the {\em isolation index\/} of the pair
$\braced{A,B}$ to be
{\small
$$
\alpha_\ssb{A,B}=
 \half\min_{a,a'\in A\atop b,b'\in B}
\braced{\max\braced{0,d(a,b)+d(a',b')-d(a,a')-d(b,b'),
d(a,b')+d(a',b)-d(a,a')-d(b,b')}}\rule{0.3in}{0in}
$$
}
\vskip 0.1in
\begin{observation}\label{obs: isolation 3 points}
$\alpha_\ssb{\braced{x},\braced{y,z}} = \frac
{d(x,y)+d(x,z)-d(y,z)}{2}$
for any three points $x,y,z$.
\end{observation}
\vskip 0.1in

\begin{figure}[ht!]
\centerline{\epsfig{file=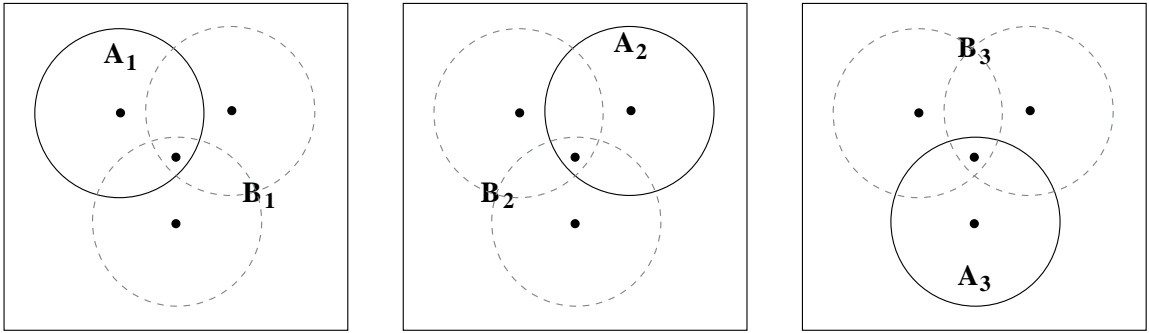, width= 6in}}
\caption{\small \bf Three splits which are not {\em weakly compatible}}
\label{fig: not compatible}
\end{figure}

A {\em split\/} of a metric
space $(X,d)$ is a partition of the points of $X$ into two non-empty
sets.
  We say that a split $A,B$ {\em separates\/} two points if one of the
points is in $A$ and the other in $B$.
We will use {\em Fraktur\/} letters for sets of splits.
If $\mathfrak{S}$ is a set of splits of $X$, we say that $\mathfrak{S}$ is
{\em weakly compatible\/} if, given any four point set $Y\subseteq X$
and given any three members of $\mathfrak{S}$, namely
 $\braced{A_1,B_1}$, $\braced{A_2,B_2}$,
and $\braced{A_3,B_3}$,
 the sets $A_1\cap Y$, $B_1\cap Y$, $A_2\cap Y$, $B_2\cap Y$,
$A_3\cap Y$, and $B_3\cap Y$ do not consist of all six two point subsets of $Y$.
Figure \ref{fig: not compatible} shows an example of three splits which are
not {\em weakly compatible}.

If $\alpha_\ssb{S} > 0$, we say that $S$ is a {\em $d$-split\/} of $X$.
The set of all $d$-splits of $X$ is always {\em weakly compatible}.
For more information regarding {\it \em weak compatibility},
see \cite{Christo97}.
From Bandelt and Dress \cite{BanDre92A},
we say that $(X,d)$ is {\em split-prime\/} if $X$ has no {\em $d$-splits\/}.
For any split $S$ the {\em split metric\/}
on $X$ is defined as
$$\delta_\ssb{S}(x,y) =
\leftbracedtwo{1 \spacedromn{if} S \spacedromn{separates}
\braced{x,y}} {0 \spacedromn{otherwise}}$$
The {\em split decomposition\/} of $(X,d)$ is defined to be
$$d = d_\ssb{0} + \sum_{S\in\mathfrak{S}}\alpha_\ssb{S}\delta_\ssb{S}$$
where $\mathfrak{S}$ is the set of all $d$-splits of $X$, and
$d_\ssb{0}$ is called the {\em split-prime residue\/} of $d$.
The {\it\em split decomposition\/} of $d$ is unique.
If $d_\ssb{0} = 0$, we say that $d$ is {\em totally decomposable}.
From Bandelt and Dress \cite{BanDre92A}, we have
\vskip 0.1in
\begin{lemma}\label{lem: 4 dec}
Every metric on four or fewer points is {\em totally decomposable\/}.
\end{lemma}
\vskip 0.1in
\begin{observation}\label{cor: factor distance}
If $X$ is totally decomposable and $x,y\in X$, then
$$d(x,y) =
 \sum_{S\spacedromn{\tiny separates}\atop\braced{x,y}}
\alpha_\ssb{S}
\delta_\ssb{S}$$
\end{observation}
\vskip 0.1in

\begin{figure}[ht!]
\centerline{\epsfig{file=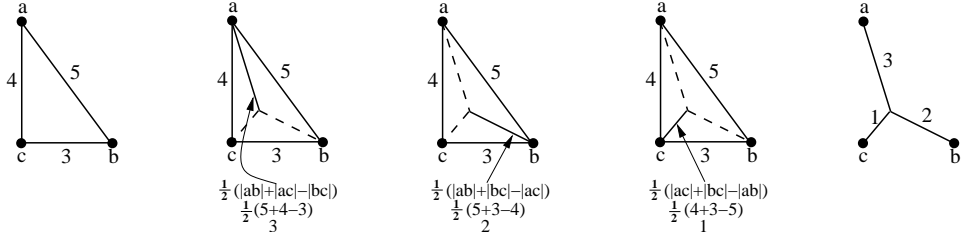, width= 5in}}
\caption{\small \bf
{\bf Step-by-step calculation of the tight span of the 3-4-5 triangle}}
\label{fig: smTS}
\end{figure}

More generally:
\vskip 0.1in
\begin{observation}\label{cor: factor distance residue}
If $d_\ssb{0}$ is the split-prime residue of $X$, and if $x,y\in X$, then
$$d(x,y) = d_\ssb{0}(x,y)+
 \sum_{S\spacedromn{\tiny separates}\atop\braced{x,y}}
\alpha_\ssb{S}
\delta_\ssb{S}$$
\end{observation}
In Figure \ref{fig: smTS}, we show how the computations of all
$d$-splits for the 3-4-5 triangle and the resulting tight span of
that space.

\subsection{\bf T-Theory and Trees}\label{subsec: new ttheory tree}

The original inspiration for the study of {\it\em T-theory\/} was the problem of
measuring how ``close" a given metric space is to being embeddable into a tree.
This question is important in {\em phylogenetic analysis}, the analysis of
relations among species or languages \cite{BarGue91,SemSte03}, since
we would like to map any set of species or languages onto a
 {\em phylogenetic tree\/} which represents their actual descent,
using a metric which represents the difference between any two members of
the set.
We say that a metric space $M$ is a {\em tree\/} if, given any two
points $x,y\in M$, there is a unique embedding of an interval of
length $d(x,y)$ into $M$ which maps the endpoints of the interval to
$x$ and $y$.  An arbitrary metric space $M$ embeds in a tree (equivalently,
$T(M)$ is a tree) if and only if $M$ satisfies the
 {\em four point condition\/} \cite{Dress84}:
\begin{eqnarray}
d(u,v)+d(x,y)&\le&\max\braced{d(u,x)+d(v,y),d(u,y)+d(v,x)}
\spacedromn{\ \ \ \ For any} u,v,x,y\in M\label{eqn: four point}
\end{eqnarray}

\subsection{\bf Tight Spans of Finite Metric Spaces}\label{subsec: new finite}

Two metric spaces $X_1$ and $X_2$ of the same cardinality
are {\em combinatorially equivalent\/} if the tight spans
$T(X_1)$ and $T(X_2)$ are combinatorially isomorphic cell complexes.
A finite metric space of cardinality $n$ is defined 
to be {\em generic\/} if $P(X)$ is
a simple polytope, \ie, if every vertex of $T(X)$ is the intersection
of exactly $n$ of the bounding hyperplanes of $P(X)$.
Equivalently, $(X,d)$ is generic
if there is some $\varepsilon > 0$ such that
$T(X,d)$ is combinatorially equivalent to $T(X,d')$ for any other metric
$d'$ on $X$ which is within $\varepsilon$ of $d$, \ie, if
$\barred{d(x,y)-d'(x,y)}<\varepsilon$ for all $x,y\in X$.

\begin{figure}[ht!]
\centerline{\epsfig{file=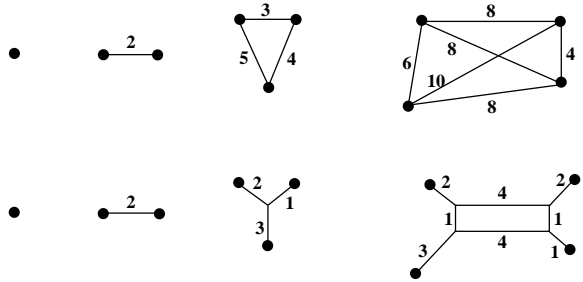, width= 3in}}
\caption{\small \bf Examples of the decomposition given by Observation
\protect\ref{cor: factor distance}
of metrics on four or fewer points.
In the upper figures, distances between points are shown.
The lower figures show the tight spans, where the edge lengths are
isolation indices.}
\label{fig: 12345 part 1}
\end{figure}

\begin{figure}[ht!]
\centerline{\epsfig{file=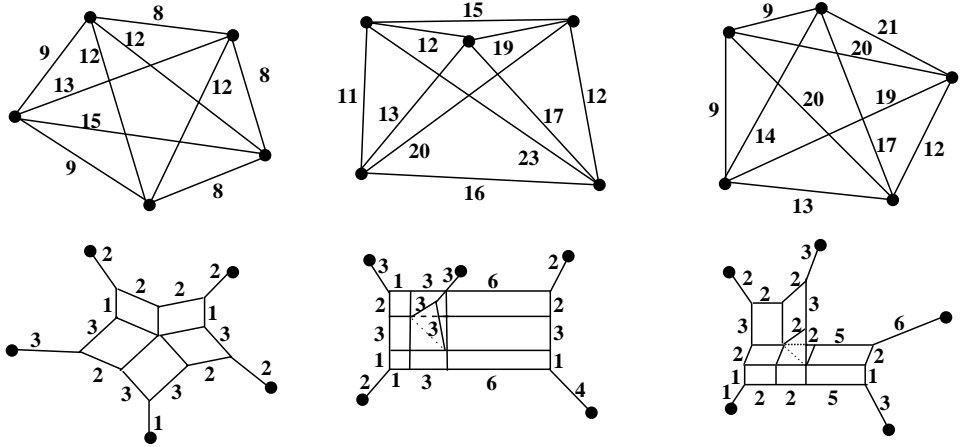, width= 5in}}
\caption{\small \bf Examples of {\it\em tight spans} for the three generic
cases of spaces with five points.
Observation \protect\ref{cor: factor distance} applies only to the first case;
Observation \protect\ref{cor: factor distance residue} applies to all cases.
}
\label{fig: 12345 part 2}
\end{figure}

The number of combinatorial classes of generic metric spaces of cardinality
$n$ increases rapidly with $n$.
There is just one combinatorial class of generic metrics for each $n \le 4$.
The tight span of one example for each $n \le 4$ is
illustrated in Figure \ref{fig: 12345 part 1}.
For $n=5$, there are three combinatorial classes of generic metrics.
One example of the tight span for each such class is illustrated in
Figure \ref{fig: 12345 part 2}.
There are 339 combinatorial classes of generic metrics for $n=6$,
as computed by Sturmfels and Yu \cite{StuYu04}.

\subsection{\bf Motivation for Using {\em T-Theory\/}
for the $k$-{\em Server Problem}}
\label{subsec: ttheory server}

\begin{figure}[ht!]
\centerline{\epsfig{file=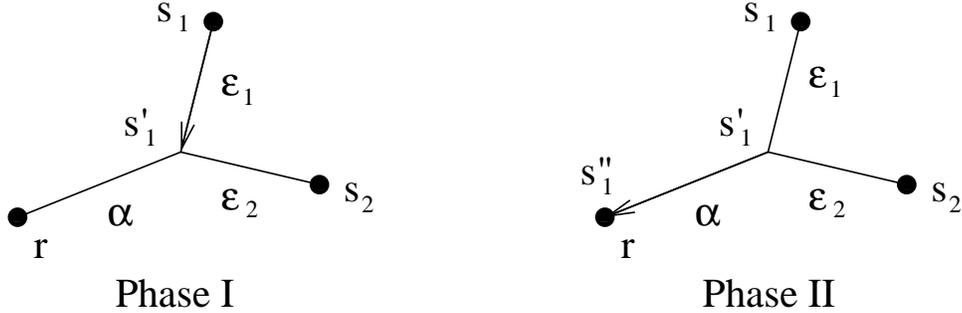, width= 5in}}
\caption{\small \bf The movement phases of $s_\ssb{1}$ to $r$}
\label{fig: motivation}
\end{figure}

In  Figure \ref{fig: motivation}, we illustrate the motivation, in the
case $k=2$, behind using T-theory to analyze the server problem.
Let
$\varepsilon_\ssb{1} = \alpha_\ssb{\braced{s_1},\braced{s_2,r}}$,
$\varepsilon_\ssb{2} = \alpha_\ssb{\braced{s_2},\braced{s_1,r}}$,
and
$\alpha = \alpha_\ssb{\braced{r},\braced{s_1,s_2}}$.
If $s_\ssb{i}$ serves the request, the total distance it moves
is $\varepsilon_\ssb{i} + \alpha$.  We can say that $\varepsilon_\ssb{i}$
is the {\em unique portion\/} of that distance, while $\alpha$ is the
{\em common portion\/}.  When we make a decision as to which server to
move, instead of comparing the two distances to $r$, we could
compare the unique portions of those distances.
In  Figure \ref{fig: motivation}, we assume that $s_\ssb{1}$ serves
the request at $r$.  The movement of $s_\ssb{1}$ can be thought
of as consisting of two {\em phases\/}.  During the first phase,
$s_\ssb{1}$ moves towards both points $r$ and $s_\ssb{2}$.
In the second phase, further movement towards both $r$ and $s_\ssb{2}$
is impossible, so $s_\ssb{1}$ moves towards $r$ and away from $s_\ssb{2}$.

In the case that $k=2$, this intuition leads to modification
of the Irani-Rubinfeld algorithm, {\sc balance2},
 \cite{IraRub91} to {\sc balance slack},
which we discuss in
\S \ref{subsec: new balance slack}, and
modification of {\sc harmonic} \cite{RagSni89} to {\sc random slack},
which we discuss in
\S \ref{subsec: new random slack}.
For $k > 3$, the intuition is
still present, but it is far less clear how to modify {\sc balance2}
and {\sc harmonic} to improve their competitivenesses.
Teia \cite{Teia93b} has partially succeeded; his algorithm
{\sc handicap}, discussed in this paper in Section \ref{sec: new handicap},
is a generalization of {\sc balance slack} to all $k$.
{\sc handicap} is trackless, and
is $k$-competitive against the lazy adversary for all $k$.
Teia \cite{Teia93b} also proves that,
for $k=3$, {\sc handicap} is 157-competitive against any adversary
(Theorem \ref{thm: handicap 157} of this paper).

\section{\bf The Virtual Server Construction}\label{sec: new virtual}

In an arbitrary metric space $M$,
the points to which we would like to move the servers may not exist.
We overcome that restriction by allowing servers to
{\em virtually\/} move in $T(M)$, while leaving the real servers in $M$.
(In an implementation, the algorithm keeps the positions of the
virtual servers in memory.)

More generally,
if $M\subseteq M'$ are metric spaces and there is a $C$-competitive
{\em online algorithm\/} $\calA'$ for the
$k$-{\em server problem\/} in $M'$, there is a
$C$-competitive {\em online algorithm\/} $\calA$
for the $k$-{\em server problem\/} in $M$.
If $\calA'$ is deterministic or randomized,
$\calA$ is deterministic or randomized, respectively.
As requests are made, $\calA$ makes use of $\calA'$ to
calculate the positions the servers of $\calA'$, which we call
{\em virtual servers}.  When there is a request $r\in M$, $\calA$
calculates the response of $\calA'$ and, in its memory, moves the
{\it\em virtual servers\/} in $M'$.
If the $i^\tH$ {\it\em virtual server\/} serves the request, then
$\calA$ moves the $i^\tH$ real server in $M$ to $r$ to serve the request,
but does not move any other real servers.

We give a formal description of the construction of $\calA$ from
$\calA'$:
\vskip 0.2in
\noindent
\begin{minipage}{6.4in}
\hrule
\vskip 0.2in
\noindent
\underline{\bf Virtual Server Construction}
\vskip 0.1in
{\small 
\begin{list}{}
\item
Let $\braced{s_\ssb{i}}$ be the servers in $M$, and let
$\braced{s'_\ssb{i}}$ be the virtual servers in $M'$.
\item
Let $s'_\ssb{i} = s_\ssb{i}$ for all $i$.
\item
Initialize $\calA'$.
\item
For each request $r$:
\begin{list}{}
\item
Move the virtual servers in $M'$ according to the algorithm $\calA'$.
\item
At least one virtual server will reach $r$.  If $s'_\ssb{i}$ reaches $r$,
move $s_\ssb{i}$ to $r$.
All other servers remain in their previous positions.
\end{list}
\end{list}
}
\vskip 0.2in
\hrule
\end{minipage}
\vskip 0.2in

We can assume that the {\it\em virtual servers\/} match the real servers
initially.
If a server $s_\ssb{i}$ serves request $r^{t}$ and then also serves
request $r^{t'}$, for some $t' > t$, then $s_\ssb{i}$ does not move
during any intermediate step.  The corresponding virtual server
can make several moves between those steps, matching the real server
at steps $t$ and $t'$.  Thus, by the triangle inequality, the movement
of each virtual server is as least as great as the movement of the
corresponding real server.
Thus, $\cost_\ssb{\calA} \le \cost_{\ssb{\calA'}}$
for the entire request sequence.
It follows that the competitiveness of $\calA$
cannot exceed the the competitiveness of $\calA'$.

\section{\bf Tree Algorithms}\label{sec: new tree}

The {\em tree algorithm}, which we call {\sc tree},
a $k$-competitive online algorithm for the $k$-server
problem in a tree, occupies
a central place in the construction of a number of the online algorithms for
the $k$-server problem presented in this paper.
The {\em line algorithm\/}, {\sc Double Coverage}, given
in \S \ref{subsec: new dc} below, is the direct ancestor
of {\sc tree}.

\subsection{\bf Double Coverage}\label{subsec: new dc}

In \cite{ChKaPV91}, Chrobak, Karloff, Payne, and Viswanathan
defined a deterministic memoryless fast
$k$-competitive {\it\em online algorithm\/}, called {\sc double coverage} (DC),
for the real line.  If a request $r$ is to the left or right of all servers,
the nearest server serves.  If $r$ is between two servers, they both
move toward the $r$ at the same speed and stop when one of them reaches $r$.

\vskip 0.2in
\noindent
\begin{minipage}{6.4in}
\hrule
\vskip 0.2in
\noindent
\underline
{\bf Double Coverage}
\vskip 0.10in
{\small
\begin{list}{}
\item
For each request $r$:
\begin{list}{}
\item
If $r$ is at the location of some server, serve $r$ at no cost.
\item
If $r$ is to the left of all servers, move the leftmost server to $r$.
\item
If $r$ is to the right of all servers, move the rightmost server to $r$.
\item
If $s_\ssb{i} < r < s_\ssb{j}$ and there are no servers in the open interval
$\parend{s_\ssb{i},s_\ssb{j}}$,
 let $\delta=\min\braced{r-s_\ssb{i},s_\ssb{j}-r}$.
  Move $s_\ssb{i}$ to the right by $\delta$,
 and move $s_\ssb{j}$ to the left by $\delta$.
At least one of those two servers will reach $r$.
\end{list}
\end{list}
}
\vskip 0.2in
\end{minipage}
\hrule
\vskip 0.2in

\begin{figure}[ht!]
\centerline{\epsfig{file=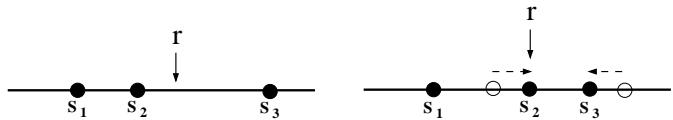, width= 3.5in}}
\caption{\small \bf The DOUBLE COVERAGE algorithm}
\label{fig: dc}
\end{figure}

\subsection{\bf The Tree Algorithm}\label{subsec: new tree}

DC is generalized by Chrobak and Larmore
in \cite{ChrLar91A} to a deterministic memoryless fast
$k$-competitive {\it\em online algorithm\/},
 {\sc tree}, for the $k$-{\it\em server problem\/} in a tree.
We can then extend {\sc tree} to any metric space which embeds in a tree,
using the virtual server construction given in Section \ref{sec: new virtual}.
\vskip 0.2in

\noindent
\begin{minipage}{6.4in}
\hrule
\nopagebreak
\vskip 0.2in
\noindent
\underline{\bf The Tree Algorithm}
\vskip 0.10in
{\small
\begin{list}{}
\item
Repeat the following loop until some server reaches $r$:
\begin{list}{}
\item
Define each server $s_\ssb{i}$ to be {\em blocked} if there is some
server $s_\ssb{j}$ such that
$d(s_\ssb{i},r) = d(s_\ssb{i},s_\ssb{j}) + d(s_\ssb{j},r)$,
and either $d(s_\ssb{j},r) < d(s_\ssb{i},r)$ or $j < i$.
Any server that is not blocked is {\em active\/}.
\item
For each $i\ne j$,
let $\alpha_\ssb{i,j} = \alpha_\ssb{\braced{s_i},\braced{s_j,r}}
=\half\parend{d(s_\ssb{i},r)+d(s_\ssb{i},s_\ssb{j})-d(s_\ssb{j},r)}$.
\item
If there is only one active server, move it to $r$.
\item
If there is more than one active server:
\begin{list}{}
\item
Let $\delta$ be the minimum value of all $\alpha_\ssb{i,j}$ for all
choices of $i,j$ such that both $s_\ssb{i}$ and $s_\ssb{j}$ are active.
\item
Move each active server a distance of $\delta$ toward $r$.
\end{list}
\end{list}
\end{list}
}
\vskip 0.2in
\hrule
\end{minipage}
\vskip 0.2in

\begin{figure}[ht!]
\centerline{\epsfig{file=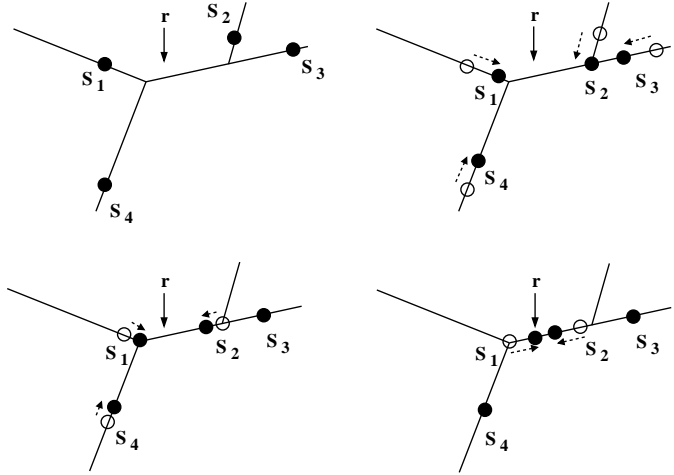, width= 3.5in}}
\caption{\small \bf The phases of one step of TREE}
\label{fig: tree}
\end{figure}

Assume $M$ is a tree.
If $s_\ssb{1}, \ldots,s_\ssb{k}$ are the servers
and $r$ is a request, we say that $s_\ssb{i}$ is {\em blocked by $s_\ssb{j}$\/}
if $d(s_\ssb{i},r) = d(s_\ssb{i},s_\ssb{j}) + d(s_\ssb{j},r)$,
and either $d(s_\ssb{j},r) < d(s_\ssb{i},r)$ or $j < i$.
Any server that is not blocked by another server is
{\em active\/}.  The algorithm serves the request by moving the servers
in a sequence of {\em phases\/}.  During each phase, all active servers
move the same distance towards $r$.  A phase ends when either one server
reaches $r$ or some previously active server becomes blocked.  After at
most $k$ phases, some server reaches $r$ and serves the request.
Figure \ref{fig: tree} illustrates an example step
(consisting of three phases) of {\sc tree} where $k=4$.
The proof of $k$-competitiveness of {\sc tree} makes
use of the {\em Coppersmith-Doyle-Raghavan-Snir potential\/}
\cite{CoDoRS93}, namely
$$\Phi_{CDRS} = \sum_{1\le i<j\le k}d(s_\ssb{i},s_\ssb{j})
+2\sum_{1\le i\le k}d(s_\ssb{i},a_\ssb{i})$$
where $\braced{a_\ssb{1},\ldots,a_\ssb{k}}$ is the set of positions of
the optimal servers and $\braced{s_\ssb{i}\leftrightarrow a_\ssb{i}}$
is the minimum matching of the algorithm servers with the optimal servers.
We refer the reader to \cite{ChrLar91A} for details of the proof.

More generally,
if $M$ satisfies the four point condition given in Inequality
(\ref{eqn: four point}), then $T(M)$ is a tree.
We simply use the above algorithm on $T(M)$ to define a
$k$-{\it\em competitive algorithm \/} on $M$, using the method of
Section \ref{sec: new virtual}.
We remark that in the original paper describing {\sc tree}
 \cite{ChrLar91A}, there was no
mention of the {\it\em tight span construction}.  The result was simply stated
using the clause, ``If $M$ embeds in a tree \ldots ."

\subsection{\bf The Slack Coverage Algorithm}\label{subsec: new bartal}

Bartal's {\em Slack Coverage algorithm\/} (SC) is
3-competitive for the 2-server problem in any Euclidean
space\footnote{A parametrized class of {\em Slack Coverage} algorithms
is described in Borodin and El-Yaniv \cite{BorElY98}.
Our definition of SC agrees with the case that the parameter is $\half$.}
 \cite{Bartal94b}.

\vskip 0.2in
\noindent
\begin{minipage}{6.4in}
\hrule
\vskip 0.2in
\noindent
\underline{Slack Coverage}
\vskip 0.1in
{\small
\begin{list}{}
\item
For each request $r$:
\begin{list}{}
\item
Without loss of generality,
$d(s_\ssb{1},r)\le d(s_\ssb{2},r)$.
\item Let $\delta = \alpha_\ssb{\braced{s_1},\braced{s_2,r}}=
\half\parend{d(s_\ssb{1},r)+d(s_\ssb{1},s_\ssb{2})-d(s_\ssb{2},r)}$.
\item Move $s_\ssb{1}$ to $r$.
\item Move $s_\ssb{2}$
 a distance of $\delta$ along a straight line toward $r$.
\end{list}
\end{list}
}
\vskip 0.2in
\hrule
\end{minipage}
\vskip 0.2in

\begin{figure}[ht!]
\centerline{\epsfig{file=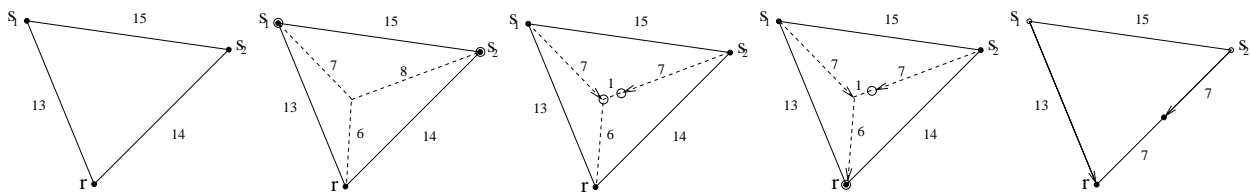, width= 6.5in}}
\caption{\small \bf One step of Bartal's Slack Coverage algorithm
in a Euclidean space: $s_\ssb{1}$ serves, and $s_\ssb{2}$
moves $\alpha_\ssb{\braced{s_1},\braced{s_2,r}} = 7$ towards $r$}
\label{fig: bartal}
\end{figure}

The intuition behind SC is that a Euclidean space $E$ is close to being
injective.  Figure \ref{fig: bartal} illustrates one step of SC.
First, construct
$T(X)$, where $X=\braced{s_\ssb{1},s_\ssb{2},r}$.  In $X$, the response
of the algorithm {\sc tree} would be to move $s_\ssb{1}$ to $r$ to serve
the request, and to move $s_\ssb{2}$ a distance of
$\delta= \alpha_\ssb{\braced{s_1},\braced{s_2,r}}$ towards $r$, in $T(X)$.
SC approximates that move by moving
$s_\ssb{2}$ that same distance in $E$ towards $r$.
We refer the reader to
pages 159--160 of \cite{BorElY98} for the proof that
SC is 3-competitive.\footnote{The {\em slack\/} is defined to be the
isolation index in \cite{ChrLar91C}, while in
 \cite{BorElY98}, {\em slack} is defined to be twice the isolation index.}

\section{\bf Balance Algorithms}\label{sec: new balance}

Informally, we say that a server algorithm is a {\em balance\/} algorithm
if it attempts, in some way, to balance the work among the servers.
Three algorithms discussed in this paper
satisfy that definition: {\sc balance2},
{\sc balance slack}, and {\sc handicap}.  

\subsection{\bf BALANCE2}\label{subsec: new balance2}

The {\em Irani-Rubinfeld algorithm\/}, also called {\sc balance2}
 \cite{IraRub91},
tries to equalize the total movement of each server.  More specifically,
when there is a request $r$, {\sc balance2} chooses to move that server
$s_\ssb{i}$ which minimizes $C_i+2\cdot d(s_\ssb{i},r)$, where $C_i$ is the
total cost incurred by $s_\ssb{i}$ on all previous moves.
{\sc balance2} is trackless and needs $O(k)$ memory.

\vskip 0.2in
\noindent
\begin{minipage}{6.4in}
\hrule
\vskip 0.2in
\noindent
\underline{BALANCE2}
{\small
\begin{list}{}
\item
Let $C_i=0$ for all $i$
\item
For each request $r$:
\begin{list}{}
\item
Pick the $i$ which minimizes $C_i+2\cdot d(s_\ssb{j},r)$.
\item
$C_i = C_i + d(s_\ssb{i},r)$.
\item
Move $s_\ssb{i}$ to $r$.
\end{list}
\end{list}
}
\vskip 0.2in
\hrule
\end{minipage}
\vskip 0.2in

From \cite{ChrLar91C} and \cite{IraRub91} we have
\vskip 0.1in
\begin{theorem}
The competitiveness of {\sc balance2} for $k=2$ is at least 6 and at most 10.
\end{theorem}
\vskip 0.1in
\noindent
The competitiveness of {\sc balance2} for $k > 2$ is open.

\subsection{\bf BALANCE SLACK}\label{subsec: new balance slack}

\begin{figure}[ht!]
\centerline{\epsfig{file=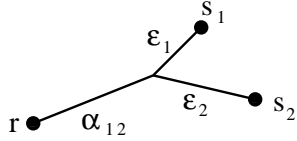, width= 1.5in}}
\caption{\small \bf Computing the moves for BALANCE SLACK and
RANDOM SLACK}
\label{fig: slack}
\end{figure}

{\sc balance slack} \cite{ChrLar91C}, defined only for $k=2$,
is a modification of {\sc balance2}.
This algorithm tries to
equalize the total {\em slack work\/}; namely, the sum, over all
requests, of the {\em Phase I\/} costs, as
illustrated in Figure \ref{fig: motivation}.

\vskip 0.2in
\noindent
\begin{minipage}{6.4in}
\hrule
\vskip 0.2in
\noindent
\underline{BALANCE SLACK}
\vskip 0.1in
{\small
\begin{list}{}
\item
Let $e_i = 0$ for $i=1,2$.
\item
For each request $r$:
\begin{list}{}
\item
$\varepsilon_\ssb{1} = \alpha_{\braced{s_\ssb{1}},\braced{s_\ssb{2},r}}$
\item
$\varepsilon_\ssb{2} = \alpha_{\braced{s_\ssb{2}},\braced{s_\ssb{1},r}}$
\item
Pick that $i$ which minimizes $e_\ssb{i}+\varepsilon_\ssb{i}$
\item
$e_\ssb{i}=e_\ssb{i}+\varepsilon_\ssb{i}$
\item
Move $s_\ssb{i}$ to $r$.
\end{list}
\end{list}
}
\vskip 0.2in
\hrule
\end{minipage}
\vskip 0.2in

We associate
each $s_\ssb{i}$ with a number $e_\ssb{i}$, the {\em slack work\/},
which is updated at each move.  If $r$ is the request point,
let $X=\braced{s_\ssb{1},s_\ssb{2},r}$, a 3-point subspace of $M$.  Let
 $\varepsilon_\ssb{1} = \alpha_{\braced{s_\ssb{1}},\braced{s_\ssb{2},r}}$
and $\varepsilon_\ssb{2} =
\alpha_{\braced{s_\ssb{2}},\braced{s_\ssb{1},r}}$,
 as shown in Figure \ref{fig: slack}.
We now update the {\em slack work values\/} as follows.  If $s_\ssb{i}$ serves
the request, we increment $e_\ssb{i}$ by adding $\varepsilon_\ssb{i}$,
while the other {\it\em slack work} remains the same.
We call $\varepsilon_\ssb{i}$ the {\em slack cost\/} of
the move if $s_\ssb{i}$ serves the request.
The algorithm {\sc balance slack} then makes that
choice which minimizes the value of $\max\braced{e_\ssb{1},e_\ssb{2}}$
after the move.

{\sc balance slack} is {\it\em trackless\/}, because it makes no use of any
information regarding any point other than the distances between the
three {\em active\/} points, namely the points of $X$, but it is not
quite {\it\em memoryless\/},
as it needs to remember one number,\footnote{It is incorrectly stated
on page 179 of \cite{BorElY98}
 that {\sc balance slack} requires unbounded memory.}
\viz, $e_\ssb{1}-e_\ssb{2}$.

From \cite{ChrLar91C} we have
\vskip 0.1in
\begin{theorem}
{\sc balance slack} is 4-competitive for $k=2$.
\end{theorem}

\subsection{\bf HANDICAP}\label{subsec: new handicap}

Teia's algorithm, {\sc handicap} \cite{Teia93b}, is also a balance algorithm,
a rather sophisticated generalization of {\sc balance slack}.
{\sc handicap} is defined for all $k$ and all metric spaces, and
is $k$-competitive against the lazy adversary.
We postpone discussion of {\sc handicap} until Section \ref{sec: new handicap}.

\section{\bf Server Algorithms in the Tight
 Span}\label{sec: new server tight span}

The tight span algorithm, {\sc tree}, and {\sc equipoise}
\cite{ChrLar91B,ChrLar91A,ChrLar92C}
permit movement of {\em virtual servers\/} in the {\it\em tight span\/}
of the metric space.
The purpose of using the tight span is that an algorithm might need to move
servers to {\it\em virtual\/}
points that do not exist in the original metric space.
The tight span, due to the {\it\em universal property\/} described in
\S \ref{subsec: new injective}, contains every virtual point
that might be needed and no others.

\subsection{\bf Virtual Servers in the
Tight Span}\label{subsec: new virtual tight}

\begin{figure}[ht!]
\centerline{\epsfig{file=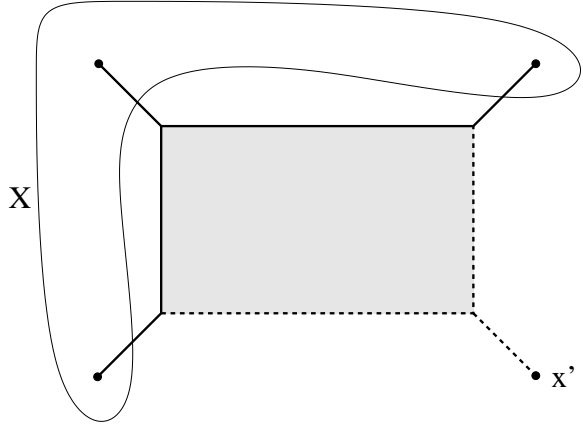, width= 3in}}
\caption{\small \bf The inclusion $\iota:T(X)\subseteq T(X')$
where $\barred{X}=3$ and $X' = X\cup\braced{x'}$}
\label{fig: iota}
\end{figure}

The tight span algorithm, {\sc tree}, and {\sc equipoise}
\cite{ChrLar91B,ChrLar91A,ChrLar92C},
 described in this paper in 
Sections \ref{sec: new tree} and \ref{sec: new server tight span}, are derived
using the embedding $M\subseteq T(M)$, from algorithms defined on $T(M)$.
One problem with that derivation is that, in the worst case, $O(\barred{M})$
numbers are required to encode a point in $T(M)$, which is impossible
if $M$ is infinite.
Fortunately, we can shortcut the process by assuming the
{\it\em virtual servers\/} are in the {\it\em tight span\/} of a finite space.
If $X\subseteq X'$ are metric spaces and $X'=X\cup\braced{x'}$,
there is a {\em canonical embedding\/}
\mbox{$\iota:T(X)\subseteq T(X')$}
 where, for any $f\in T(X)$:

$$(\iota(f))(x) = \leftbracedtwo{f(x)\spacedromn{if}x\in X}
{\sup_{y\in X}\braced{d(x',y)-f(y)}\spacedromn{if}x=x'}$$
By an abuse of notation, we identify $f$ with $\iota(f)$.
In Figure \ref{fig: iota}, $X$ consists of three points, and
$T(X)$ is the union of the solid line segments, while $T(X')$ is
the entire figure, where $X'=X\cup\braced{x'}$.

Continuing with the construction, let
$s^\malyzero_\ssb{1}, \ldots, s^\malyzero_k\in M$ be the
initial positions of the servers, and $r^\malyone \ldots r^n$ the
request sequence.  Let $X^t = \braced{s^\malyzero_\ssb{1},
\ldots,s^\malyzero_\ssb{k},r^1, \ldots,r^t}$, a set
of cardinality of at most $k+t$, for $0\le t\le n$.
Before the $t^\tH$ request all {\it\em virtual
servers\/} are in $T(X^{t-1})$.

Let $\calA'$ be an online algorithm for the $k$-server problem in
$T(M)$ and $\calA$ the algorithm in $M$ derived from $\calA'$
using the virtual server construction of Section \ref{sec: new virtual}.
When the request $r\ssp{t}$ is received,
$\calA$ uses the canonical embedding $T(X^{t-1})\subseteq T(X^t)$
 to calculate the positions of
the {\it\em virtual servers\/} in $T(X^t)$, then uses $\calA'$ to move the
{\it\em virtual servers\/} within $T(X^t)$.
At most, $\calA$ is required to remember the distance of each virtual
server to each point in $X^t$.

\subsection{\bf The Tight Span Algorithm}\label{subsec: new tight span}

{\sc tree} of \S \ref{subsec: new tree} generalizes to
all metric spaces in the case that $k=2$, essentially because $T(X)$
is a tree for any metric space $X$ with at most three points.
This generalization was first defined in \cite{ChrLar91A}, but was not
named in that paper.  We shall call it the {\em tight span algorithm}.
As we did for {\sc tree}, we first define the {\it\em tight span} algorithm
as a fast memoryless algorithm in any injective metric
space.
We then use the virtual construction of Section \ref{sec: new virtual}
to extend the definition of the tight span algorithm to any metric space.

\vskip 0.2in
\noindent
\begin{minipage}{6.4in}
\hrule
\nopagebreak
\vskip 0.2in
\noindent
\underline{\bf The Tight Span Algorithm}
{\small
\vskip 0.1in
\begin{list}{}
\item For each request $r$:
\begin{list}{}
\item
Let $X = \braced{s_\ssb{1},s_\ssb{2},r}$.
\item
Pick an embedding $T(X)\subseteq M$.
\item
Execute {\sc tree} on $T(X)$.
\end{list}
\end{list}
}
\vskip 0.2in
\nopagebreak
\hrule
\end{minipage}
\vskip 0.2in

\begin{figure}[ht!]
\centerline{\epsfig{file=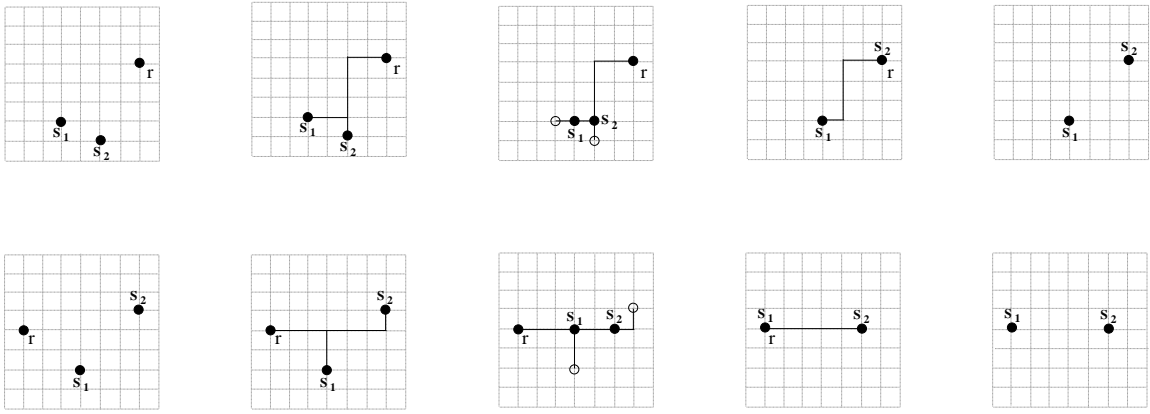, width= 6in}}
\caption{\small \bf The {\em tight span algorithm\/} in the
 {\it\em Manhattan plane}}
\label{fig: manhattan}
\end{figure}

Assume that $M$ is injective, \ie, $M=T(M)$.
We define the {\em tight span algorithm\/} on $M$ as follows:
let  $X=\braced{s_\ssb{1}, s_\ssb{2},r}\subseteq M$.
Since $M$ is injective, the inclusion $X\subseteq M$
can be extended to an embedding of $T(X)$ into $M$.
Since $T(X)$ is a tree, use {\sc tree}
to move both servers in $T(X)$ such
that one of the servers moves to $r$.  Since $T(X)\subseteq M$,
we can move the servers in $M$.
In Figure \ref{fig: manhattan}, we show an example consisting of
two steps of the
{\it\em tight span algorithm}, where $M$ is the {\it\em Manhattan plane}.

Finally, we extend the {\it\em tight span algorithm\/} to an arbitrary metric
space by using the virtual server
construction given in Section \ref{sec: new virtual}.
We refer the reader to \cite{ChrLar91B}
for the proof of 2-{\it\em competitiveness\/} for $k=2$, which also uses the
Coppersmith-Doyle-Raghavan-Snir potential.

\subsection{\bf EQUIPOISE}\label{subsec: new equipoise}

 \begin{figure}[ht!]
 \centerline{\epsfig{file=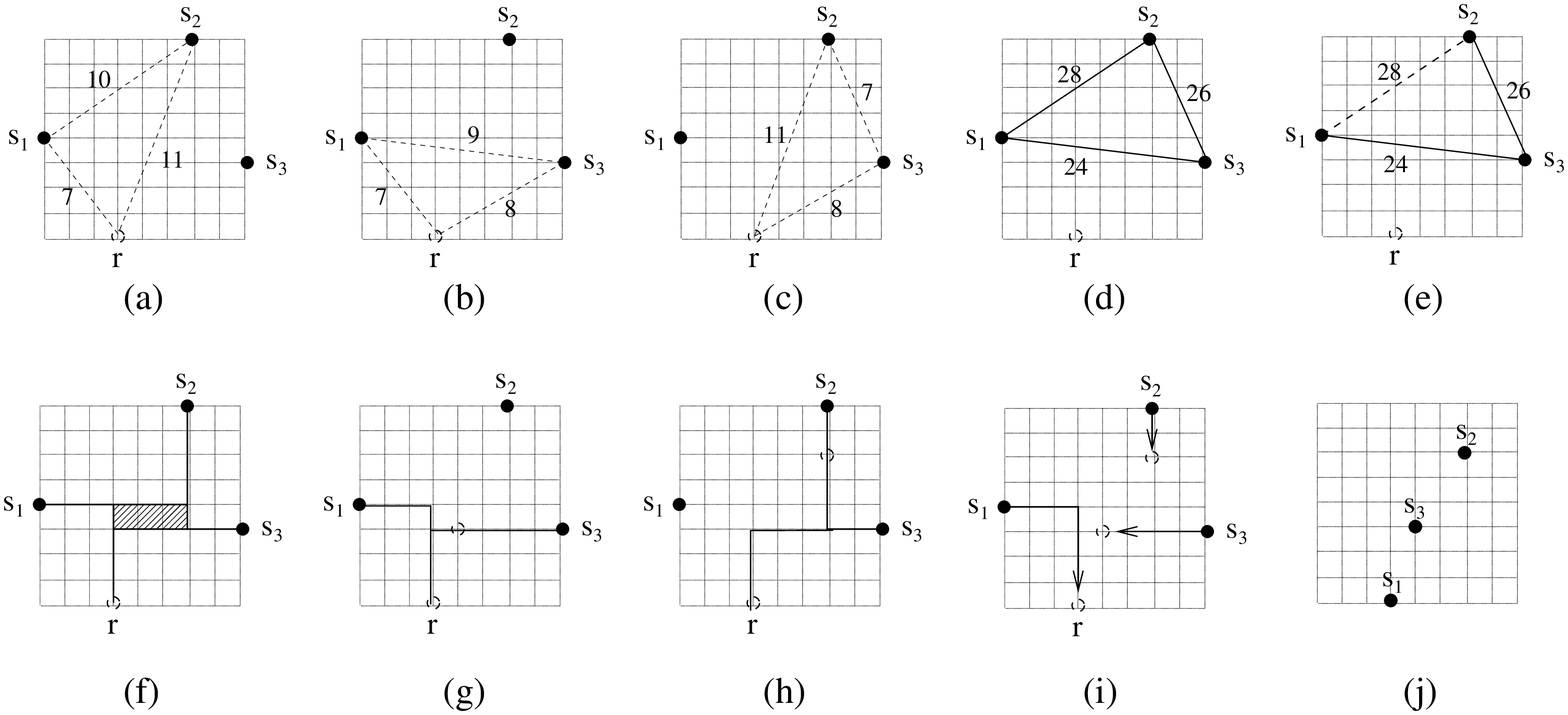, width= 6in}}
 \caption{\small \bf One step of EQUIPOISE in the {\it\em Manhattan plane}
\newline
 \noindent
$\bullet$
 Figure \ref{fig: equipoise}(a) shows the computation of $w_\ssb{12}$.
\newline
 \noindent
$\bullet$
 Figure \ref{fig: equipoise}(b) shows the computation of $w_\ssb{13}$.
\newline
 \noindent
$\bullet$
 Figure \ref{fig: equipoise}(c) shows the computation of $w_\ssb{23}$.
\newline
 \noindent
$\bullet$
 Figure \ref{fig: equipoise}(d) shows the weighted graph $G$.
\newline
 \noindent
$\bullet$
 Figure \ref{fig: equipoise}(e) indicates
 $E_{MST}= \braced{e_\ssb{13},e_\ssb{23}}$, the minimum spanning tree of $G$.
\newline
 \noindent
$\bullet$
 Figure \ref{fig: equipoise}(f) shows $T(X)$, with
   the two-dimensional cell of $T(X)$ shaded.
\newline
 \noindent
$\bullet$
 Figure \ref{fig: equipoise}(g) shows $T_{e_{13}}$, and the position
 $s_\ssb{3}$ would move to if {\sc tree} for two servers were executed on
 $T_{e_{13}}$
\newline
 \noindent
$\bullet$
 Figure \ref{fig: equipoise}(h) shows $T_{e_{23}}$, and the position
 $s_\ssb{2}$ would move to if {\sc tree} for two servers were executed on
 $T_{e_{23}}$.
\newline
 \noindent
$\bullet$
 Figure \ref{fig: equipoise}(i) shows the minimum matching movement of
 $S$ to $S'$.
\newline
 \noindent
$\bullet$
 Figure \ref{fig: equipoise}(j) shows the positions of the three servers
 after completion of the step.
 }
 \label{fig: equipoise}
 \end{figure}

In \cite{ChrLar92C}, a {\it\em deterministic algorithm\/}
for the $k$-{\em server problem\/},
called {\sc equipoise}, is given.  For $k=2$, {\sc equipoise} is
the {\em tight span algorithm\/} of \cite{ChrLar91B}
discussed in \S \ref{subsec: new tight span}, and is 2-competitive.
For $k=3$, {\sc equipoise} is 11-competitive.
The competitiveness of {\sc equipoise} for $k\ge 4$ is unknown.

\vskip 0.2in
\noindent
\begin{minipage}{6.4in}
\hrule
\nopagebreak
\vskip 0.2in
\nopagebreak
\noindent
\underline{\bf EQUIPOISE}
\vskip 0.1in
{\small
\begin{list}{}
\item For each request $r$:
\begin{list}{}
\item
Let $G$ be the complete graph whose nodes are
$S=\braced{s_\ssb{1},\ldots,s_\ssb{k}}$\\ and whose
edges are $E=\braced{e_\ssb{i,j}}$.
\item
For each $1\le i < j \le k$, let
$w_\ssb{i,j} = d(s_\ssb{i},s_\ssb{j})+d(s_\ssb{i},r)+d(s_\ssb{j},r)$
be the {\em weight\/} of $e_\ssb{i,j}$.
\item
Let $E_{MST}\subseteq E$ be the edges of a minimum spanning tree of $G$.
\item
For each $e=e_\ssb{i,j}\in E_{MST}$:
\begin{list}{}
\item
Let $T_e$ be the tight span of $\braced{s_\ssb{i},s_\ssb{j},r}$.
Choose an embedding $T_e\subseteq M$.
\item
Emulate {\sc tree} on $T_e$ for two servers at $s_\ssb{i}$ and
$s_\ssb{j}$ and request point $r$. One of those servers will move
to $r$, while the other will move to some point $p_e\in T_e\subseteq M$.
\end{list}
\item
Let $S'=\braced{r}\cup\braced{p_e\ |\ e\in E_\ssb{MST}}$, a set of
cardinality $k$.
\item
Move the servers to $S'$, using a minimum matching of $S$
and $S'$.  One server will move to $r$.
\end{list}
\end{list}
}
\vskip 0.2in
\hrule
\end{minipage}
\vskip 0.2in

Let $M$ be an arbitrary metric space.
We first define {\sc equipoise} assuming that $M$ is injective.
Let $S = \braced{s_\ssb{1},\ldots,s_\ssb{k}}$, the configuration of our
servers in $M$, let $r$ be the request point,
and let $X=\braced{s_\ssb{1}, \ldots,s_\ssb{k},r}$.
Let $G$ be the complete weighted graph whose vertices are $S$ and whose
edge weights are $\braced{w_{i,j}}$, where
$w_{i,j} = d(s_\ssb{i},s_\ssb{j})+d(s_\ssb{i},r)+d(s_\ssb{j},r)$
for any $i\ne j$,
  Let $E_\ssb{MST}$
be the set of edges of a minimum spanning tree for $G$.

For each $e=e_\ssb{i,j} = \braced{s_\ssb{i},s_\ssb{j}}\in E_\ssb{MST}$,
let $X_{e}=\braced{s_\ssb{i},s_\ssb{j},r}$, and let $T_e=T(X_e)$,
and choose an embedding $T_e\subseteq M$.
We then use the algorithm {\sc tree}, for two servers, as a subroutine.
For each $e=e_\ssb{i,j}\in E_{MST}$, we consider how {\sc tree}
would serve the request $r$
if its two servers were at $s_\ssb{i}$ and $s_\ssb{j}$.  It would move
one of those servers to $r$, and the other to some other point in $M$,
which we call $p_e$.
Let $S'=\braced{r}\cup\braced{p_e\ |\ e\in E_\ssb{MST}}$,
a set of cardinality $k$.
{\sc equipoise} then serves the request at $r$ by moving its servers
from $S$ to $S'$, using the minimum matching of those two sets.
One server will move to $r$, serving the request.
Figure \ref{fig: equipoise} shows one step of {\sc equipoise}
in the case $k=3$, where $M$ is the {\em Manhattan plane}.
By using the {\em virtual server\/} construction of
Section \ref{sec: new virtual}, we extend {\sc equipoise} to all metric spaces.

\section{\bf Definition and Analysis of HANDICAP}\label{sec: new handicap}

In this section, we define the algorithm {\sc handicap}, a generalization
of {\sc balance slack}, given initially
in Teia's dissertation \cite{Teia93b}, using slightly different notation.
{\sc handicap} is trackless and fast, but not memoryless.

The algorithm given in \cite{ChrLar91D} is $k$-competitive against the
lazy adversary, but only if the adversay is {\em benevolent\/} \ie, informs
us when our servers matches his; {\sc handicap} is more general, since
it does not have that restriction.

For $k\le 3$, {\sc handicap} has the least competitiveness of any known
deterministic trackless algorithm for the $k$-server problem.
The competitiveness of {\sc handicap} for $k\ge 4$ is unknown.

We give a proof that, for all $k$,
{\sc handicap} is $k$-competitive against the lazy
adversary; in fact, against any adversary that can have at most one
open server, \ie, a server in a position different from any of the
algorithm's servers.  This result was proved in \cite{Teia93b}.
The proof given here is a
simplification inspired by Teia \cite{Teia93d}.

\subsection{\bf Definition of HANDICAP}\label{subsec: new define handicap}

{\sc handicap} maintains numbers $E_1, \ldots, E_k$, where $E_i$
is called the {\em handicap\/}\footnote{In
 \cite{Teia93b}, the handicap was defined to be $H_i$.  The value
of $H_i$ is twice $E_i$.}
of the $i^\tH$ server.
The handicap of each server is updated after every step, and is used
to decide which server moves.  The larger a server's handicap, the
less likely it is to move.
Since only the differences of the handicaps are used,
the algorithm remembers only $k-1$ numbers between steps.

\vskip 0.2in
\noindent
\begin{minipage}{6.4in}
\hrule
\vskip 0.2in
\noindent
\underline{\bf HANDICAP}
\vskip 0.1in
{\small
\begin{list}{}
\item
Let $E_j = 0$ for all $j$
\item
For each request $r$:
\begin{list}{}
\item
Pick that $i$ for which $E_i + d(s_\ssb{i},r)$ is minimized.
\item
For all $1\le j\le k$:\\
\rule{0.2in}{0in}$E_j = E_j +
 \alpha_{\braced{r},\braced{s_\ssb{i},s_\ssb{j}}}$
\item
Move $s_\ssb{i}$ to $r$.
\end{list}
\end{list}
}
\vskip 0.2in
\hrule
\end{minipage}
\vskip 0.2in

Initially, all handicaps are zero.
At any step,
let $s_\ssb{1}, \ldots, s_\ssb{k}$ be the positions of our servers, and
let $r$ be the request point.  For all $1\le i,j\le k$, define
$\alpha_{_{ij}} = \alpha_{_{\{r\},\{s_\ssb{i},s_\ssb{j}\}}}$,
the isolation index.
Choose that $i$ for which $E_i+d(s_\ssb{i},r)$ is minimized, breaking ties
arbitrarily.  Next, update the handicaps by adding $\alpha_{_{ij}}$
to $E_j$ for each $j$, and then move the $i^\tH$ server to $r$.
The other servers do not move.
It is a simple exercise to prove, for $k=2$,
 that {\sc handicap} $\equiv$ {\sc balance slack}.  Simply verify that
$e_\ssb{1}-e_\ssb{2} = E_1-E_2$, and that
$e_\ssb{1} + \varepsilon_\ssb{1} \le e_\ssb{2} + \varepsilon_\ssb{2}$
if and only if $E_1+d(s_\ssb{1},r)\le E_2+d(s_\ssb{2},r)$.

Let
$a_\ssb{1}, \ldots, a_\ssb{k}$ be the adversary's servers.
We assume
that the indices are assigned in such a way that
$\braced{ s_{_\ssb{i}}\leftrightarrow a_{_\ssb{i}}}$ is a minimum matching.
If $s_\ssb{i}\ne a_\ssb{i}$,
we say that $s_\ssb{i}\leftrightarrow a_\ssb{i}$ is an {\em open matching\/},
and $a_\ssb{i}$ is an {\em open server\/}.
If $s_\ssb{i} = a_\ssb{i}$ for all $i$,
we can arbitrarily designate any $a_\ssb{i}$ to be the open server.
We now prove that {\sc handicap} is $k$-competitive against any
adversary which may not have more than one open server, using
the {\em Teia potential\/} defined below, a simplification
of the potential used in \cite{Teia93b}.

\begin{figure}[ht!]
\centerline{\epsfig{file=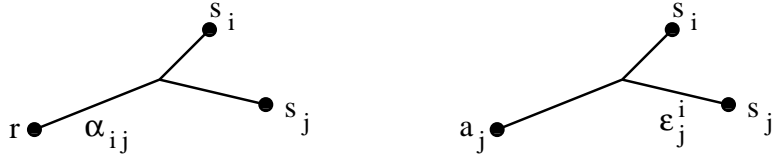, width= 4in}}
\caption{\small \bf Definitions of $\alpha_{ij}$ and
 $\varepsilon^\ssp{i}_\ssb{j}$ for {\sc handicap}}
\label{fig: handicap alpha epsilon}
\end{figure}

\subsection{\bf Competitiveness of HANDICAP Against the Lazy
Adversary}\label{subsec: new proof handicap}

In order to aid the reader's intuition,
we define the {\em Teia potential $\Phi$},
to be a sum of simpler quantities.
\begin{center}
$
\begin{array}{rcl}
\romn{\bf Name}&
\romn{\bf Notation}&
\romn{{\bf Formula}}\\
\rule{0in}{0.2in}
\romn{Server Diversity}&
\calD
&
\sum\limits_{1\le i < j\le k}d(
s_\ssb{i},
s_\ssb{j})\\
\rule{0in}{0.3in}
\romn{$k\cdot$(Minimum Matching)}&
\calM
&
k\cdot
\sum\limits_{i=1}^kd(
a_\ssb{i},
s_\ssb{i})\\
\rule{0in}{0.2in}
\romn{
Coppersmith-Doyle-Raghavan-Snir Potential
}&
\Phi_{CDRS}&\calD+\calM\\
\rule{0in}{0.2in}
\romn{Tension ({\em Spannung}) Induced by $s_\ssb{i}$ on
$\braced{a_\ssb{j},s_\ssb{j}}$}&
\varepsilon^i_\ssb{j}
&
\alpha_{\{s_\ssb{j}\},\{a_\ssb{j},s_\ssb{i}\}}
 \spacedromn{for all}i,j\\
\rule{0in}{0.3in}
\romn{Total Tension Induced by $s_\ssb{i}$}&
\varepsilon^i
&
\sum\limits_{j=1}^k\varepsilon^i_\ssb{j}
\spacedromn{for all}i\\
\rule{0in}{0.2in}
\romn{Net Handicap of $s_\ssb{i}$}&
e_\ssb{i}
&
E_i-\varepsilon^i
\spacedromn{for all}i\\
\rule{0in}{0.2in}
\romn{Maximum Net Handicap}&
e_{\max}
&
\max\limits_\ssb{1\le i\le k}
e_\ssb{i}\\
\rule{0in}{0.3in}
\romn{Handicap Portion of Potential}&
\calH & 2\cdot
\sum\limits_{i=1}^k
\parend{
e_{\max}
-E_i
}
\\
\rule{0in}{0.2in}
\romn{Teia Potential}&
\Phi
&
\Phi_{CDRS}+
\calH
\end{array}
$
\end{center}
\vskip 0.1in

We prove that $\Phi$ is non-negative, and that
the following {\em update condition\/} holds for each step:
\begin{eqnarray}
\Delta\Phi-k\cdot\cost_\ssb{\adv}+\cost_\ssb{\alg} &\le& 0
\label{eqn: update handicap}
\end{eqnarray}
where $\cost_\ssb{\alg}$ and $\cost_\ssb{\adv}$
are the algorithm's and the adversary's costs for the step, and
where $\Delta\Phi$ is the change in potential during that step.

\begin{lemma}\label{lem: phi nonnegative}
$\Phi\ge 0$.
\end{lemma}

\begin{proof} By all $i$:
\begin{eqnarray*}
 2\varepsilon^i_\ssb{1}
&=&
d(s_\ssb{1},s_\ssb{i})+d(s_\ssb{1},a_\ssb{1})-d(a_\ssb{1},s_\ssb{i})
\hspace{0.5in}\spacedromn{by Observation \ref{obs: isolation 3 points}}
\\
&\le&
d(s_\ssb{1},s_\ssb{i})+d(s_\ssb{1},a_\ssb{1})
\\
\spacedromn{Thus}\rule{1.2in}{0in}&&\\
\Phi
&=&
\calD+\calM+\calH\\
&=&
\sum\limits_{1\le i < j\le k}d(
s_\ssb{i},
s_\ssb{j})
+
\sum\limits_{1\le i < j\le k}d(
s_\ssb{i},
s_\ssb{j})
+
2\cdot
\sum\limits_{i=1}^k
\parend{
e_{\max}
-E_i
}
\\
&\ge&
\sum\limits_{i=1}^k
d(s_i,s_1)
+k\cdot
d(a_1,s_1)
+
2\cdot
\sum\limits_{i=1}^k
\parend{
e_{\max}
-E_i
}
\\
&=&
\sum^k_{i=1}\parend{d(s_\ssb{1},s_\ssb{i})+d(s_\ssb{1},a_\ssb{1})
+2\parend{e_{_{\max}} - e_\ssb{i}-\varepsilon^i_\ssb{1}}}\\
\\
&=&
\sum^k_{i=1}\parend{\parend{d(s_\ssb{1},s_\ssb{i})+d(s_\ssb{1},a_\ssb{1})
-2\varepsilon^i_\ssb{1}}
+2\parend{e_{_{\max}} - e_\ssb{i}}}\\
&\ge&
0
\end{eqnarray*}
\end{proof}

Every step can be factored into a combination of
two kinds of moves:
\begin{enumerate}
\item The adversary can move its open server to some other point, but
make no request.  We call this a {\em cryptic move\/}.
\item The adversary can request the position of its open server,
without moving any server.  We call this a {\em lazy request\/}.
\end{enumerate}
To prove that Inequality (\ref{eqn: update handicap}),
the {\em update condition},
holds for every step, it suffices to prove that it holds for every
cryptic move and for every lazy request.

\begin{lemma}\label{lem: update handicap cryptic}
Inequality\/ {\rm (\ref{eqn: update handicap})}
 holds for a cryptic move.
\end{lemma}
\begin{proof}
We use the traditional $\Delta$ notation throughout
to indicate the increase of any quantity.

Without loss of generality, $r=a_\ssb{1}$, the open server.
Let $\hat a_\ssb{1}$ be the new position of
the adversary's server.
Then $\Delta\calD = 0$, since the positions of the algorithm's servers
do not change, $\Delta\calM = k\cdot\parend{
 d(s_\ssb{1},\hat a_\ssb{1})-d(s_\ssb{1}, a_\ssb{1})}$,
and \mbox{$\Delta e_\ssb{i} = -\Delta\varepsilon_\ssb{1}^i$}
 for each $i$.
Since $\Delta e_\ssb{\max}\le \max_i\Delta e_\ssb{i}$, there exists some
$j$ such that
\begin{eqnarray*}
\Delta\calH
&\le& - 2k\cdot \Delta \varepsilon_\ssb{1}^j\\
&=&
k\parend{d(s_\ssb{1},a_\ssb{1})-d(a_\ssb{1},s_\ssb{j})
-d(s_\ssb{1},\hat a_\ssb{1})+d(\hat a_\ssb{1},s_\ssb{j})}
\ \ \romn{by Observation \ref{obs: isolation 3 points}}\\
\spacedromn{Thus}\rule{1.2in}{0in}&&\\
\Delta\Phi-k\cdot\cost_\ssb{\adv}
&=&
\Delta\calM+\Delta\calH-k\cdot d(a_\ssb{1},\hat a_\ssb{1})\\
&\le&
k\parend{d(\hat a_\ssb{1},s_\ssb{j})
-d(a_\ssb{1},s_\ssb{j})- d(a_\ssb{1},\hat a_\ssb{1})}
\\
&\le& 0
\end{eqnarray*}
\end{proof} 

\vskip 0.2in
\begin{lemma}\label{lem: update handicap lazy}
Inequality\/ {\rm (\ref{eqn: update handicap})}
 holds for a lazy request.
\end{lemma}
\begin{proof} 
Without loss of generality,
 $a_\ssb{1}$
is the open server.  Then $r=a_\ssb{1}$, the request point.

\begin{figure}[ht!]
\centerline{\epsfig{file=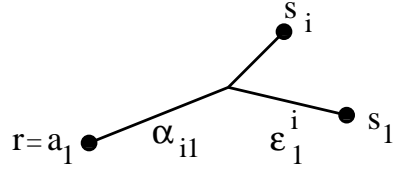, width= 2in}}
\caption{\small \bf Illustration of the proof of Lemma
\protect\ref{lem: update handicap lazy}}
\label{fig: handicap move open}
\end{figure}

{\bf Case I:} $s_\ssb{1}$ serves the request.

\noindent
As illustrated in Figure \ref{fig: handicap move open},
$ \alpha_{\ssb{i}\ssb{1}} + \varepsilon^i_\ssb{1}
= d(s_\ssb{1}, a_\ssb{1})$, for all $i$,
$\Delta\calD =
\sum^k_{i=2}\parend{\alpha_{\ssb{i}\ssb{1}} - \varepsilon^i_\ssb{1}}$,
and \mbox{$\alpha_{\ssb{1}\ssb{1}} = d(s_\ssb{1}, a_\ssb{1})$}.
Then
\begin{eqnarray*}
\Delta E_i \hspace{0.12in}
&=&
\alpha_{\ssb{i}\ssb{1}}
\rule{1.32in}{0in}
\spacedromn{for all}i
\\
\Delta e_\ssb{i} \hspace{0.15in}
&=&
\alpha_{\ssb{i}\ssb{1}} + \varepsilon^i_\ssb{1} = d(s_\ssb{1},a_\ssb{1})
\rule{.26in}{0in}
\spacedromn{for all}i\\
\Delta e_\ssb{\max}
&=&
d(s_\ssb{1},a_\ssb{1})\rule{1in}{0in}
\spacedromn{since}
\min\Delta e_\ssb{i}\le \Delta e_\ssb{\max}\le \max\Delta e_\ssb{i}\\
\spacedromn{Thus}\rule{1in}{0in}&&\\
\Delta\Phi+\cost_\ssb{\alg} &=&
\Delta\calD+\Delta\calM+\Delta\calH+d(s_\ssb{1},a_\ssb{1})\\
&=&
\sum^k_{i=2}\parend{\alpha_{\ssb{i}\ssb{1}} - \varepsilon^i_\ssb{1}}
-k\cdot d(s_\ssb{1},a_\ssb{1})+2\parend{k\cdot d(s_\ssb{1},a_\ssb{1})
-\sum^k_{i=1}\alpha_{\ssb{i}\ssb{1}}}+d(s_\ssb{1},a_\ssb{1})\\
&=&
(k+1)\cdot d(s_\ssb{1},a_\ssb{1})
-\sum^k_{i=2}\parend{\alpha_{\ssb{i}\ssb{1}}+\varepsilon^i_\ssb{1}}
-2\cdot\alpha_\ssb{11}
\\
&=&0\ \ 
\parend{\spacedromn{since}
\alpha_{\ssb{i}\ssb{1}}+\varepsilon^i_\ssb{1}
=d(s_\ssb{1},a_\ssb{1})=\alpha_\ssb{11}}
\end{eqnarray*}

{\bf Case II:} For some $i > 1$, $s_\ssb{i}$ serves the request.
Without loss of generality, $i=2$.  Using the carat notation
to indicate the updated values after the move, we have
$\hat s_\ssb{2} = \hat a_\ssb{2} = a_\ssb{1}$,
$\hat s_\ssb{1} = a_\ssb{2}$,
$\hat s_\ssb{i}=s_\ssb{i}$ for all $i > 2$, and
$\hat a_\ssb{i}=a_\ssb{i}$ for all $i\ne 2$.

\vskip 0.1in
{\bf Claim A:} $\Delta e_\ssb{i} = \alpha_\ssb{12}$ for all $i\ne 2$.

\noindent
Using Observation \ref{obs: isolation 3 points}:
\begin{eqnarray*}
e_\ssb{i} 
&=&
E_i-\varepsilon^i_\ssb{1}\\
\hat e_\ssb{i}
&=&
E_i+\alpha_\ssb{2i}-\hat\varepsilon^i_\ssb{1}\\
\varepsilon^i_\ssb{1}
&=&
\frac
{
d(s_\ssb{i},s_\ssb{1})
+d(s_\ssb{1},a_\ssb{1})
-d(s_\ssb{i},a_\ssb{1})
}{2}\\
\alpha_\ssb{2i}
&=&
\frac
{
d(s_\ssb{2},a_\ssb{1})
+d(s_\ssb{i},a_\ssb{1})
-d(s_\ssb{2},s_\ssb{i})}{2}
\\
\hat \varepsilon^i_\ssb{1}
&=&
\frac
{
d(s_\ssb{i},s_\ssb{1})
+d(s_\ssb{1},a_\ssb{2})
-d(s_\ssb{i},a_\ssb{2})
}{2}
\\
\alpha_\ssb{12}
&=&
\frac
{
d(s_\ssb{1},a_\ssb{1})
+d(s_\ssb{2},a_\ssb{1})
-d(s_\ssb{1},s_\ssb{2})
}{2}
\end{eqnarray*}
Combining the above equations, we obtain
$\hat e_\ssb{i} - e_\ssb{i} - \alpha_\ssb{12} = 0$,
which verifies Claim A.

\vskip 0.1in
{\bf Claim B:} $\hat e_\ssb{\max} = \hat e_\ssb{i}$ for some $i\ne 2$.

\noindent
Since {\sc handicap} moves $s_\ssb{2}$, we know that
$E_2+d(s_\ssb{2},a_\ssb{1})\le E_1 + d(s_\ssb{1},a_\ssb{1})$.
\begin{eqnarray*}
\hat e_\ssb{1} &=&
 \hat E_1 + \alpha_\ssb{12} =
 E_1 + \alpha_\ssb{12}\rule{1in}{0in}
 \spacedromn{by Claim A}\\
\hat e_\ssb{2} &=&
 \hat E_2 - \hat\varepsilon^\malytwo_\ssb{1} =
 E_2 + d(s_\ssb{2},a_\ssb{1}) - \hat\varepsilon^\malytwo_\ssb{1}\\
\spacedromn{Thus}\rule{1in}{0in}&&\\
\hat e_\ssb{1}-\hat e_\ssb{2}
&=&
E_1 + \alpha_\ssb{12} - E_2 - d(s_\ssb{2},a_\ssb{1})
 + \hat\varepsilon^\malytwo_\ssb{1}
\rule{2in}{0in}
\\
&\ge&
\alpha_\ssb{12}- d(s_\ssb{1},a_\ssb{1}) + \hat\varepsilon^\malytwo_\ssb{1}\\
&=& 0
\end{eqnarray*}
Since $\hat e_\ssb{1} \ge \hat e_\ssb{2}$, we have verified Claim B.
\vskip 0.1in

\noindent
We now continue with the proof of Case II of
Lemma \ref{lem: update handicap lazy}.
From Claims A and B, $\Delta e_\ssb{\max} \le \alpha_\ssb{12}$.
Recall that $s_\ssb{2}=a_\ssb{2}$, and
$r=a_\ssb{1}$.  Thus
$\alpha_\ssb{22} =
\alpha_{_{\{r\},\{s_\ssb{2},s_\ssb{2}\}}}= d(s_\ssb{2},a_\ssb{1})$.
   Then
\begin{eqnarray*}
\Delta\calD
&=&
\sum_{i\ne 2}\parend{
d(s_\ssb{i},a_\ssb{1})-
d(s_\ssb{i},s_\ssb{2})
}
\\
\Delta\calM
&=&
k\parend{
d(s_\ssb{1},a_\ssb{2})-
d(s_\ssb{1},a_\ssb{1})
} =
k\parend{
d(s_\ssb{1},s_\ssb{2})-
d(s_\ssb{1},a_\ssb{1})
}
\\
\Delta\calH
&\le&
2k\cdot\alpha_\ssb{12}
-2\sum_{i=1}^k\alpha_\ssb{2i}\\
\Delta\Phi+\cost_\ssb{\alg}
&=&
\Delta\calD+\Delta\calM+\Delta\calH+d(s_\ssb{2},a_\ssb{1})\\
&\le&
\sum_{i\ne 2}\parend{
d(s_\ssb{i},a_\ssb{1})-
d(s_\ssb{i},s_\ssb{2})
}
+
k\parend{
d(s_\ssb{1},s_\ssb{2})-
d(s_\ssb{1},a_\ssb{1})
}
+
2k\cdot\alpha_\ssb{12}
\\
&&
\ \ \ \ -2\sum_{i=1}^k\alpha_\ssb{2i}
+d(s_\ssb{2},a_\ssb{1})\\
&=&
k\parend{2\alpha_\ssb{12}-d(s_\ssb{1},a_\ssb{1})+d(s_\ssb{1},s_\ssb{2})}
\\
&&
\ \ \ \ -\sum_{i\ne 2}\parend{
\alpha_\ssb{2i}-d(s_\ssb{i},a_\ssb{1})+d(s_\ssb{i},s_\ssb{2})}
-2\alpha_\ssb{22}
+d(s_\ssb{2},a_\ssb{1})\\
&=&
kd(s_\ssb{2},a_\ssb{1})-(k-1)d(s_\ssb{2},a_\ssb{1})
-2d(s_\ssb{2},a_\ssb{1}) +d(s_\ssb{2},a_\ssb{1})
\\
&=&
0
\end{eqnarray*}
\vskip 0.1in
\noindent
This completes the proof of Lemma \ref{lem: update handicap lazy},
since the left-hand side of the update condition is less than or equal to
zero.
\end{proof} 
\vskip 0.1in
\noindent
\begin{theorem}\label{thm: handicap}
{\sc handicap} is $k$-competitive against any adversary which can have
at most one open server.
\end{theorem}
\begin{proof}
Lemma \ref{lem: phi nonnegative} states that the Teia potential is
non-negative, while Lemmas 
\ref{lem: update handicap cryptic} and \ref{lem: update handicap lazy}
state that the update condition, Inequality (\ref{eqn: update handicap}),
holds for every step.
\end{proof}

\vskip 0.1in
\noindent
Teia also obtains a competitiveness of {\sc handicap} against any
adversary, for $k=3$.
From Subsection 8.4 of Teia's dissertation \cite{Teia93b}, on page 59:
\vskip 0.1in
\begin{theorem}\label{thm: handicap 157}
For $k=3$, {\sc handicap} is 157-competitive.
\end{theorem}

\section{\bf Harmonic Algorithms}\label{sec: new harmonic}

In this section, we present the classical algorithm {\sc harmonic},
as well as {\sc random slack}, an improvement of {\sc harmonic}
which uses T-theory.
In \S \ref{subsec: new harmonic analysis} we present a T-theory based
proof that {\sc harmonic} is 3-competitive for $k=2$.

\subsection{\bf HARMONIC}\label{subsec: new harmonic}

{\sc harmonic} is a {\em memoryless randomized algorithm\/} for the
$k$-{\em server problem}, first defined by Raghavan and Snir
\cite{RagSni89,RagSni94}.
{\sc harmonic} is based on the intuition that it should be less likely
to move a larger distance than a smaller. {\sc harmonic} moves
each server with a probability that is inversely proportional to its
distance to the request point.

\vskip 0.2in
\noindent
\begin{minipage}{6.4in}
\hrule
\vskip 0.2in
\noindent
\underline{\bf HARMONIC}
\vskip 0.1in
{\small
\begin{list}{}
\item
For each request $r$:
\begin{list}{}
\item
For each $1\le i\le k$, let
\begin{eqnarray}
p_\ssb{i}
&=&
\frac{\frac{1}{d(s_\ssb{i},r)}}{\frac{1}{d(s_\ssb{1},r)}+\cdots
+\frac{1}{d(s_\ssb{k},r)}}\rule{1.2in}{0in}
\label{eqn: harmonic probability}
\end{eqnarray}
\item
Pick one $i$, where each $i$ is picked with probability $p_i$.
\item
Move $s_\ssb{i}$ to $r$.
\end{list}
\end{list}
}
\vskip 0.2in
\hrule
\end{minipage}
\vskip 0.2in

{\sc harmonic} is known to be 3-competitive for $k=2$
\cite{ChrLar92A,ChrSga00a}.
Raghavan and Snir \cite{RagSni89,RagSni94} 
 prove that its competitiveness
cannot be less than $\parend{k+1\atop 2}$, which is greater
than the best known deterministic competitiveness of the $k$-server problem
 \cite{KouPap94A,KouPap95A}.
For $k > 2$, the true competitiveness of {\sc harmonic}
is unknown but finite \cite{Grove91}.
{\sc harmonic} is of interest because it is simple to implement.

\subsection{\bf RANDOM SLACK}\label{subsec: new random slack}

 {\sc random slack\/}, defined only for two servers,
is derived from {\sc harmonic}, but
moves each server with a
probability inversely proportional to the
{\em unique\/} distance that a server would move to serve the request,
namely the {\em Phase I\/} cost (see Figure \ref{fig: motivation}.)

\vskip 0.2in
\noindent
\begin{minipage}{6.4in}
\hrule
\vskip 0.2in
\noindent
\underline{\bf RANDOM SLACK}
\vskip 0.1in
{\small
\begin{list}{}
\item For each request $r$:
\begin{list}{}
\item
$\varepsilon_\ssb{1} = \alpha_{\braced{s_\ssb{1}},\braced{s_\ssb{2},r}}$
\item
$\varepsilon_\ssb{2} = \alpha_{\braced{s_\ssb{2}},\braced{s_\ssb{1},r}}$
\item
Let $p_\ssb{1} = \frac{\varepsilon_\ssb{1}}
{\varepsilon_\ssb{1}+\varepsilon_\ssb{2}}$
\item
Let $p_\ssb{2} = \frac{\varepsilon_\ssb{2}}
{\varepsilon_\ssb{1}+\varepsilon_\ssb{2}}$
\item
Pick one $i$, where each $i$ is picked with probability $p_i$.
\item
Move $s_\ssb{i}$ to $r$.
\end{list}
\end{list}
}
\vskip 0.2in
\hrule
\end{minipage}
\vskip 0.2in

We refer the reader to \cite{ChrLar91C} for the proof that {\sc random slack}
is 2-competitive.

\subsection{\bf Analysis of HARMONIC using Isolation
Indices}\label{subsec: new harmonic analysis}

The original proof that {\sc harmonic} is 3-competitive for $k=2$
used {\it\em T-theory\/}, but was never published.
In this section, we present an updated version of that unpublished proof.

We will first show that the {\it\em lazy
potential\/} $\Phi$, defined below,
satisfies an {\em update condition\/} for every possible move.
In a manner similar to that in the proof of Theorem \ref{thm: handicap},
we first factor all moves into three kinds,
which we call {\em active\/}, {\em lazy\/}, and {\em cryptic\/}.
After each step, {\sc harmonic}'s two servers are located at points
$s_\ssb{1}$ and $s_\ssb{2}$, and the adversary's servers are located
at points $a_\ssb{1}$ and $a_\ssb{2}$.
Without loss of generality, $s_\ssb{2} = a_\ssb{2}$ is the last
request point.   In the next step, the adversary moves a server to a point
$r$ and makes a request at $r$, and then
{\sc harmonic} moves one of its two servers to $r$,
using the probability distribution given in Equation
(\ref{eqn: harmonic probability}).
We analyze the problem by requiring that the adversary always do one
of three things:
\begin{enumerate}
\item Move the server at $a_\ssb{2}$ to a new point $r$, and then
request $r$.  We call this an {\em active request\/}.
\item Request $a_\ssb{1}$ without moving a server.
We call this a {\em lazy request\/}.
\item Move the server from $a_\ssb{1}$ to some other point, but
make no request.  We call this a {\em cryptic move\/}.
\end{enumerate}
If the adversary moves its server from $a_\ssb{1}$ to a new point $r$
and then requests $r$, we consider that step to consist of two moves:
a {\it\em cryptic move\/} followed by a {\it\em lazy request\/}.
Our analysis will be simplified by this factorization.

If $x,y,z \in M$, we define $\Phi(x,y,z)$, the {\em lazy potential\/},
to be the expected cost that
{\sc harmonic} will pay if $x=s_\ssb{1}$, $y=a_\ssb{1}$, and
$z=s_\ssb{2}=a_\ssb{2}$, providing the adversary makes only lazy requests
henceforth.
The formula for $\Phi$ is obtained by solving
the following two simultaneous equations:
\begin{eqnarray}
\Phi(x,y,z) &=& \frac{2\cdot d(x,y)\cdot d(y,z)}{d(x,y)+d(y,z)}
 + \frac{d(x,y)}{d(x,y)+d(y,z)}\cdot \Phi(x,z,y)
\label{eqn: lazy}
\\
\Phi(x,z,y) &=& \frac{2\cdot d(x,z)\cdot d(y,z)}{d(x,z)+d(y,z)}
 + \frac{d(x,z)}{d(x,z)+d(y,z)}\cdot \Phi(x,y,z)\\
\romn{Obtaining the solution}&&\nonumber
\\
\Phi(x,y,z) &=& \frac
{2\cdot d(x,y)(2\cdot d(x,z)+d(y,z))}
{d(x,y)+d(x,z)+d(y,z)}
\end{eqnarray}

\begin{theorem}\label{thm: harmonic 3 comp}
{\sc harmonic} is 3-competitive for 2 servers.
\end{theorem}

\begin{proof}
We will show that the {\it\em lazy potential\/} is 3-competitive.
For each move, we need to verify the {\it\em update condition\/}, namely
that the value of $\Phi$ before the move, plus three times
the distance moved by the adversary server, is at least as great
as the expected distance moved by {\sc harmonic} plus the
expected value of $\Phi$ after the move.  
The update condition holds for every {\it\em lazy request\/},
by Equation (\ref{eqn: lazy}).
We need to verify the update inequalities
for {\it\em active requests\/} and {\it\em cryptic moves\/}.

If {\sc harmonic} has servers at $x$ and $z$ and the adversary has
servers at $y$ and $z$,
the update condition for the active request where the adversary
moves the server from $z$ to $r$ is:

{\footnotesize
\begin{eqnarray}
\Phi(x,y,z) + 3\cdot d(z,r)
- \frac{2\cdot d(x,r)\cdot d(z,r)}{d(x,r)\cdot d(z,r)}
- \frac{d(x,r)}{d(x,r)\cdot d(z,r)}\cdot \Phi(x,y,r)
- \frac{d(z,r)}{d(x,r)\cdot d(z,r)}\cdot \Phi(z,y,r)
 &\ge& 0
\label{harmonic update condition active}
\end{eqnarray}
}

If {\sc harmonic} has servers at $x$ and $z$ and the adversary has
servers at $y$ and $z$,
the update condition for the {\it\em cryptic move\/} where the adversary
moves the server from $y$ to $r$ is:

\begin{eqnarray}
\Phi(x,y,z) +3\cdot d(y,r) - \Phi(x,r,z)
 &\ge& 0
\label{harmonic update condition cryptic}
\end{eqnarray}

Let $X=\braced{x,y,z,r}$.
It will be convenient to choose a variable name for the
{\it\em isolation index\/} of each split of $X$.  Let:
\begin{eqnarray*}
a &=& \alpha_{\braced{x},\braced{y,z,r}}\\
b &=& \alpha_{\braced{y},\braced{x,z,r}}\\
c &=& \alpha_{\braced{z},\braced{x,y,r}}\\
d &=& \alpha_{\braced{r},\braced{x,y,z}}\\
e &=& \alpha_{\braced{x,y},\braced{z,r}}\\
f &=& \alpha_{\braced{x,z},\braced{y,r}}\\
g &=& \alpha_{\braced{x,r},\braced{y,z}}
\end{eqnarray*}
By Lemma \ref{lem: 4 dec} and Observation \ref{cor: factor distance}, we have
\begin{eqnarray*}
d(x,y) &=& a+b+f+g\\
d(x,z) &=& a+c+e+g\\
d(y,z) &=& b+c+e+f\\
d(x,r) &=& a+d+e+f\\
d(y,r) &=& b+d+e+g\\
d(z,r) &=& c+d+f+g
\end{eqnarray*}

\begin{figure}[ht!]
\centerline{\epsfig{file=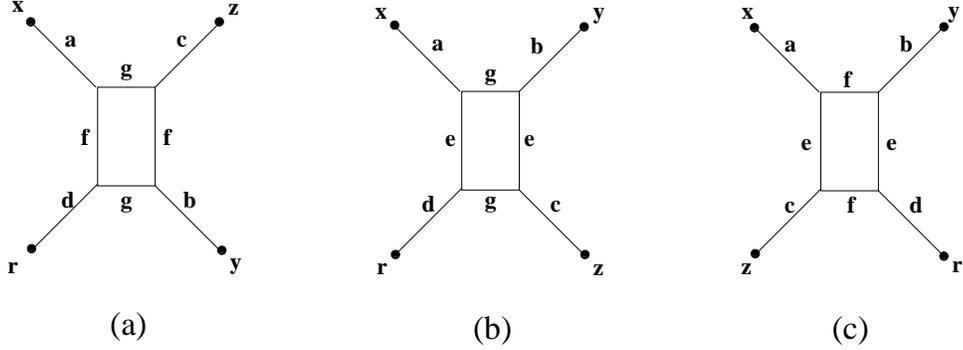, width= 5in}}
\caption{\small
 \bf Three possible pictures of the {\em tight span\/} of $X$}
\label{fig: prime harmonic}
\end{figure}

The three non-trivial splits of $X$ do not form a coherent set; thus, at least
one of their {\it\em isolation indices\/} must be zero.
Figure \ref{fig: prime harmonic} shows the three
generic possibilities for $T(X)$.
Let $MaxMatch = \max\braced{d(x,y)+d(z,r),d(x,z)+d(y,r),d(x,r)+d(y,z)}$.
If $MaxMatch = d(x,y)+d(z,r)$, then $e=0$, as shown in
Figure \ref{fig: prime harmonic}(a).
If $MaxMatch = d(x,z)+d(y,r)$, then $f=0$, as shown in
 Figure \ref{fig: prime harmonic}(b).
If $MaxMatch = d(x,r)+d(y,z)$, then $g=0$, as shown in
 Figure \ref{fig: prime harmonic}(c).
In any case, the product $efg$ must be zero.  Substituting the 
formula for each distance, and using the fact that $efg=0$, we
compute the left hand side of Inequality
(\ref{harmonic update condition active})
to be $$\frac{\italicized{numerator}_\ssb{1}}{
(a+b+c+e+f+g) (a+b+d+e+f+g) 
(b+c+d+e+f+g) (a+c+2 d+e+2 f+g)}$$
\noindent
where {\it numerator$_\ssb{1}$} is a polynomial in the literals
$a,b,c,d,e,f,$ and $g$, given in Appendix \ref{app: math}.

Similarly, the left hand side of Inequality
(\ref{harmonic update condition cryptic})
is
$$\frac{\italicized{numerator}_\ssb{2}}
{(a+b+c+e+f+g)(a+c+d+e+f+g)}$$
where {\it numerator$_\ssb{2}$} is also a polynomial in the literals
$a,b,c,d,e,f,$ and $g$, given in Appendix \ref{app: math}.

The denominators of these rational expressions are clearly positive.
The proof that $\italicized{numerator}_\ssb{1}$
and $\italicized{numerator}_\ssb{2}$
are non-negative is given in Appendix \ref{app: math}.
Thus, the left hand sides of both inequalities are non-negative,
thus verifying that {\sc harmonic} is 3-competitive for two servers.
\end{proof}

\section{\bf Summary and Possible Future Applications of T-Theory to the
k-Server Problem}\label{sec: new summary future}

We have demonstrated the usefulness of {\it\em T-theory\/} for defining
{\it\em online algorithms\/} for the {\it\em server problem\/}
in a metric space $M$, and 
proving competitiveness
by rewriting the {\it\em update conditions\/}
in terms of {\it\em isolation indices}.
In this section, we suggest ways to extend the use of T-theory to
obtain new results for the server problem.

\subsection{\bf Using {\em T-Theory\/} to Generalize
RANDOM SLACK}\label{subsec: trackless}

A {\it\em memoryless trackless randomized algorithm\/} for the
$k$-{\it\em server problem \/} must act as follows.  Given that the servers
are at
$\braced{s_\ssb{1}\ldots,s_\ssb{k}}$
 and the request is $r$, first compute $T(X)$,
where \mbox{$X = \braced{s_\ssb{1}\ldots,s_\ssb{k},r}$}
and then use the parameters of $T(X)$ to compute the probabilities
of serving the request with the various servers.

We know that this approach is guaranteed to yield a
{\it\em competitive memoryless randomized algorithm\/}
 for the $k$-{\it\em server problem\/}, since {\sc harmonic}
is in this class.  {\sc harmonic} computes probabilities
using the parameters of $T(X)$, but we saw in {\sc random slack}
in \S \ref{subsec: new random slack}
that, for $k=2$, a more careful choice of probabilities yields an improvement
of the competitiveness.  We conjecture that, for $k \ge 3$,
there is some choice of probabilities which yields an algorithm of
this class whose competitiveness is lower than that of {\sc harmonic}.

\subsection{\bf Using {\em T-Theory\/} to Analyze
HARMONIC for Larger $k$}\label{subsec: harmonic}

We know that the competitiveness of {\sc harmonic} for $k=3$ is at
least $\parend{4\atop 2} = 6$
 \cite{RagSni89,RagSni94}.
  As in Section \ref{sec: new harmonic},
we could express the {\it\em lazy potential\/} in closed form, and then
 attempt to prove that it satisfies all necessary update conditions.

In principle, the process of verifying that the {\it\em lazy potential\/}
suffices to prove 6-competitiveness for {\sc harmonic} for $k=3$
could be automated, possibly using the output of Sturmfels and Yu's
program \cite{StuYu04} as input.
However, the complexity of the proof technique used in Section
\ref{sec: new harmonic} rises very rapidly with $k$, and may be impractical
for $k > 2$.  There should be some way to simplify this computation.

\subsection{\bf Generalizing the Virtual Server
Algorithms and the $k$-Server Conjecture}\label{subsec: original}

The $k$-server conjecture remains open, despite years of effort by
many researchers.  The most promising approach to date appears to be
the effort to prove that the {\em work function algorithm\/} (WFA)
 \cite{ChrLar92B}, or perhaps a variant of WFA, is $k$-competitive.
This opinion is explained in depth by Koutsoupias \cite{Koutso94}.
To date, for $k\ge 3$, it is only known that WFA is $(2k-1)$-competitive
 \cite{KouPap95A}, and that it is $k$-competitive in a number of
special cases.

{\sc handicap} represents a somewhat different
approach to the $k$-server problem.
Teia conjectures that {\sc handicap} can be modified in such a way
as to obtain a 3-competitive deterministic online algorithm for the
3-server problem against an arbitrary adversary,
thus settling the server conjecture for $k=3$.
He suggests that this can be done by maintaining two reference points
 in the tight span.
The resulting algorithm would not be trackless.
\vskip 0.25in

\noindent
From the introduction (pp. 3--4) of Teia's dissertation
\cite{Teia93b}:

\vskip 0.25in

\begin{center}
\begin{minipage}{5.8in}
{
For the case of more than one open matching, the memory representation
would have to be augmented by additional components.  One possibility would be
to introduce reference points in addition to handicaps.  We are
convinced that, for $k = 3$, by careful case analysis and the introduction
of two reference points, a 3-competitive algorithm can be given.}
\end{minipage}
\end{center}
\vskip 0.25in
\begin{center}
\begin{minipage}{6.4in}
Original German text:
\vskip 0.25in
\begin{center}
\begin{minipage}{5.8in}
\it
F\"ur den Fall mehr als eines offenen\/ {\em Matchings} m\"u{\ss}te die
Ged\"achtnisrepr\"asentation um zus\"atzliche Komponenten erweitert werden.
Eine M\"oglichkeit w\"are, zus\"atzlich zu den\/ {\em \mbox{Handicaps}}
Bezugspunkte einzuf\"uhren.
Wir sind \"uberzeugt, da{\ss} sich f\"ur $k=3$ durch sorgf\"altige
Fallunterscheidungen und die Einf\"uhrung zweier Bezugspunkte ein
\mbox{3- kompetitiver} Algorithmus angeben l\"a{\ss}t.
\end{minipage}
\end{center}
\end{minipage}
\end{center}

\section*{Acknowledgment}
We wish to thank Dean Bartkiw and Marek Chrobak for reviewing the final
manuscript.

\bibliographystyle{plain}
\bibliography{online,ttheory,books,extra}

\begin{thebibliography}{10}

\bibitem{AlKaPW95}
Noga Alon, Richard~M. Karp, David Peleg, and Douglas West.
\newblock A graph-theoretic game and its application to the k-server problem.
\newblock {\em SIAM J. Comput.}, 24:78--100, 1995.

\bibitem{BalShe93}
Ganesh~R. Baliga and Anil~M. Shende.
\newblock On space bounded server algorithms.
\newblock In {\em Proc. 5th International Conference on Computing and
  Information}, pages 77--81. IEEE, 1993.

\bibitem{BanChe96}
Hans-J{\"u}rgen Bandelt and Victor Chepoi.
\newblock Embedding metric spaces in the rectilinear plane: a six-point
  criterion.
\newblock {\em GEOMETRY: Discrete \& Computational Geometry}, 15:107--117,
  1996.

\bibitem{BanChe98}
Hans-J{\"u}rgen Bandelt and Victor Chepoi.
\newblock Embedding into the rectilinear grid.
\newblock {\em NETWORKS: Networks: An International Journal}, 32:127--132,
  1998.

\bibitem{BanDre92A}
Hans-J{\"u}rgen Bandelt and Andreas Dress.
\newblock A canonical decomposition theory for metrics on a finite set.
\newblock {\em Adv. Math.}, 92:47--105, 1992.

\bibitem{Bartal94b}
Yair Bartal.
\newblock A fast memoryless 2-server algorithm in {E}uclidean spaces, 1994.
\newblock Unpublished manuscript.

\bibitem{BaChLa98}
Yair Bartal, Marek Chrobak, and Lawrence~L. Larmore.
\newblock A randomized algorithm for two servers on the line.
\newblock In {\em Proc. 6th European Symp. on Algorithms (ESA)}, Lecture Notes
  in Comput. Sci., pages 247--258. Springer, 1998.

\bibitem{BaChLa00}
Yair Bartal, Marek Chrobak, and Lawrence~L. Larmore.
\newblock A randomized algorithm for two servers on the line.
\newblock {\em Inform. and Comput.}, 158:53--69, 2000.

\bibitem{BaChNR02}
Yair Bartal, Marek Chrobak, John Noga, and Prabhakar Raghavan.
\newblock More on random walks, electrical networks, and the harmonic
  $k$-server algorithm.
\newblock {\em Inform. Process. Lett.}, 84:271--276, 2002.

\bibitem{BarGro00}
Yair Bartal and Edward Grove.
\newblock The harmonic $k$-server algorithm is competitive.
\newblock {\em J. ACM}, 47(1):1--15, 2000.

\bibitem{BarKou04}
Yair Bartal and Elias Koutsoupias.
\newblock On the competitive ratio of the work function algorithm for the
  $k$-server problem.
\newblock {\em Theoret. Comput. Sci.}, 324:337--345, 2004.

\bibitem{BarMen04}
Yair Bartal and Manor Mendel.
\newblock Randomized $k$-server algorithms for growth-rate bounded graphs.
\newblock In {\em Proc. 15th Symp. on Discrete Algorithms (SODA)}, pages
  666--671. ACM/SIAM, 2004.

\bibitem{BarRos92}
Yair Bartal and Adi Ros\'en.
\newblock The distributed $k$-server problem --- a competitive distributed
  translator for $k$-server algorithms.
\newblock In {\em Proc. 33rd Symp. Foundations of Computer Science (FOCS)},
  pages 344--353. IEEE, 1992.

\bibitem{BarGue91}
Jean-Pierre Barth{\'e}lemy and Alain Gu{\'e}noche.
\newblock {\em Trees and Proximity Relations,}.
\newblock Wiley, Chichester, 1991.
\newblock Translated from the French by Gregor Lawden.

\bibitem{BeChLa99}
Wolfgang Bein, Marek Chrobak, and Lawrence~L. Larmore.
\newblock The 3-server problem in the plane.
\newblock In {\em Proc. 7th European Symp. on Algorithms (ESA)}, volume 1643 of
  {\em Lecture Notes in Comput. Sci.}, pages 301--312. Springer, 1999.

\bibitem{BeChLa02}
Wolfgang Bein, Marek Chrobak, and Lawrence~L. Larmore.
\newblock The 3-server problem in the plane.
\newblock {\em Theoret. Comput. Sci.}, 287:387--391, 2002.

\bibitem{BeiLar00}
Wolfgang Bein and Lawrence~L. Larmore.
\newblock Trackless online algorithms for the server problem.
\newblock {\em Inform. Process. Lett.}, 74:73--79, 2000.

\bibitem{BeKaTa90}
Piotr Berman, Howard Karloff, and Gabor Tardos.
\newblock A competitive algorithm for three servers.
\newblock In {\em Proc. 1st Symp. on Discrete Algorithms (SODA)}, pages
  280--290. ACM/SIAM, 1990.

\bibitem{BlKaRS92}
Avrim Blum, Howard Karloff, Yuval Rabani, and Michael Saks.
\newblock A decomposition theorem and lower bounds for randomized server
  problems.
\newblock In {\em Proc. 33rd Symp. Foundations of Computer Science (FOCS)},
  pages 197--207. IEEE, 1992.

\bibitem{BlKaRS00}
Avrim Blum, Howard Karloff, Yuval Rabani, and Michael Saks.
\newblock A decomposition theorem and lower bounds for randomized server
  problems.
\newblock {\em SIAM J. Comput.}, 30:1624--1661, 2000.

\bibitem{BorElY98}
Allan Borodin and Ran El-Yaniv.
\newblock {\em Online Computation and Competitive Analysis}.
\newblock Cambridge University Press, 1998.

\bibitem{BuBuIv01}
Dmitri Burago, Yuri Burago, and Sergei Ivanov.
\newblock {\em A Course in Metric Geometry}.
\newblock AMS: Graduate Studies in Mathematics, v. 33, 2001.
\newblock ISBN 0-8218-2129-6.

\bibitem{Chepoi97}
Victor Chepoi.
\newblock A {$T_X$}-approach to some results on cuts and metrics.
\newblock {\em Advances Applied Mathematics}, 19:453--470, 1997.

\bibitem{Christo97}
George Christopher.
\newblock {\em Structure and Applications of Totally Decomposable Metrics}.
\newblock PhD thesis, Carnegie Mellon University, 1997.

\bibitem{ChKaPV91}
Marek Chrobak, Howard Karloff, Tom~H. Payne, and Sundar Vishwanathan.
\newblock New results on server problems.
\newblock {\em SIAM J. Discrete Math.}, 4:172--181, 1991.

\bibitem{ChrLar91B}
Marek Chrobak and Lawrence~L. Larmore.
\newblock A new approach to the server problem.
\newblock {\em SIAM J. Discrete Math.}, 4:323--328, 1991.

\bibitem{ChrLar91D}
Marek Chrobak and Lawrence~L. Larmore.
\newblock A note on the server problem and a benevolent adversary.
\newblock {\em Inform. Process. Lett.}, 38:173--175, 1991.

\bibitem{ChrLar91C}
Marek Chrobak and Lawrence~L. Larmore.
\newblock On fast algorithms for two servers.
\newblock {\em J. Algorithms}, 12:607--614, 1991.

\bibitem{ChrLar91A}
Marek Chrobak and Lawrence~L. Larmore.
\newblock An optimal online algorithm for $k$ servers on trees.
\newblock {\em SIAM J. Comput.}, 20:144--148, 1991.

\bibitem{ChrLar92A}
Marek Chrobak and Lawrence~L. Larmore.
\newblock {HARMONIC} is three-competitive for two servers.
\newblock {\em Theoret. Comput. Sci.}, 98:339--346, 1992.

\bibitem{ChrLar92B}
Marek Chrobak and Lawrence~L. Larmore.
\newblock The server problem and on-line games.
\newblock In Lyle~A. McGeoch and Daniel~D. Sleator, editors, {\em On-line
  Algorithms}, volume~7 of {\em DIMACS Series in Discrete Mathematics and
  Theoretical Computer Science}, pages 11--64. AMS/ACM, 1992.

\bibitem{ChrLar92C}
Marek Chrobak and Lawrence~L. Larmore.
\newblock Generosity helps or an 11-competitive algorithm for three servers.
\newblock {\em J. Algorithms}, 16:234--263, 1994.

\bibitem{ChrLar98}
Marek Chrobak and Lawrence~L. Larmore.
\newblock Metrical task systems, the server problem, and the work function
  algorithm.
\newblock In Amos Fiat and Gerhard~J. Woeginger, editors, {\em Online
  Algorithms: {The} State of the Art}, pages 74--94. Springer, 1998.

\bibitem{ChLaLR97}
Marek Chrobak, Lawrence~L. Larmore, Carsten Lund, and Nick Reingold.
\newblock A better lower bound on the competitive ratio of the randomized
  2-server problem.
\newblock {\em Inform. Process. Lett.}, 63:79--83, 1997.

\bibitem{ChrSga00a}
Marek Chrobak and Ji{\v{r}}{\'\i} Sgall.
\newblock A simple analysis of the harmonic algorithm for two servers.
\newblock {\em Inform. Process. Lett.}, 75:75--77, 2000.

\bibitem{ChrSga04}
Marek Chrobak and Ji{\v{r}}{\'\i} Sgall.
\newblock The weighted 2-server problem.
\newblock {\em Theoret. Comput. Sci.}, 324:289--312, 2004.

\bibitem{CoDoRS93}
Don Coppersmith, Peter~G. Doyle, Prabhakar Raghavan, and Marc Snir.
\newblock Random walks on weighted graphs and applications to on-line
  algorithms.
\newblock {\em J. ACM}, 40:421--453, 1993.

\bibitem{DrHuMo01}
Andreas Dress, Katharina~T. Huber, and Vincent Moulton.
\newblock Metric spaces in pure and applied mathematics.
\newblock In {\em Proceedings of the Conference on Quadratic Forms and Related
  Topis, LSU-2001}, pages 121--139. Documenta Mathematica.

\bibitem{Dress84}
Andreas W.~M. Dress.
\newblock Trees, tight extensions of metric spaces, and the cohomological
  dimension of certain groups.
\newblock {\em Advances in Mathematics}, 53:321--402, 1984.

\bibitem{Dress89}
Andreas W.~M. Dress.
\newblock Towards a classification of transitive group actions on finite metric
  spaces.
\newblock {\em Advances in Mathematics}, 74:163--189, 1989.

\bibitem{DrHuKM01}
Andreas W.~M. Dress, Katharina~T. Huber, Jacobus~H. Koolen, and Vincent
  Moulton.
\newblock Six points suffice: How to check for metric consistency.
\newblock {\em Eur. J. Comb.}, 22(4):465--474, 2001.

\bibitem{DrHuMo02}
Andreas W.~M. Dress, Katharina~T. Huber, and Vincent Moulton.
\newblock Antipodal metrics and split systems.
\newblock {\em Eur. J. Comb.}, 23(2):187--200, 2002.

\bibitem{DrHuMo96}
Andreas W.~M. Dress, Daniel~H. Huson, and Vincent Moulton.
\newblock Analyzing and visualizing sequence and distance data using
  splitstree.
\newblock {\em Discrete Applied Mathematics}, 71(1-3):95--109, 1996.

\bibitem{DrMoTe96}
Andreas W.~M. Dress, Vincent Moulton, and Werner Terhalle.
\newblock T-{T}heory: An overview.
\newblock {\em European J. Combinatorics}, 17(2-3):161--175, 1996.

\bibitem{DreSch87}
Andreas W.~M. Dress and Rudolf Scharlau.
\newblock Gated sets in metric spaces.
\newblock {\em Aequationes Math.}, 34:112--120, 1987.

\bibitem{ElYKle95}
Ran El-Yaniv and J.~Kleinberg.
\newblock Geometric two-server algorithms.
\newblock {\em Inform. Process. Lett.}, 53:355--358, 1995.

\bibitem{EpImSt03}
Leah Epstein, Csanad Imreh, and Rob van Stee.
\newblock More on weighted servers or {FIFO} is better than {LRU}.
\newblock {\em Theoret. Comput. Sci.}, 306:305--317, 2003.

\bibitem{FiRaRa94}
Amos Fiat, Yuval Rabani, and Yiftach Ravid.
\newblock Competitive k-server algorithms.
\newblock {\em J. Comput. Systems Sci.}, 48:410--428, 1994.

\bibitem{FiRaRS94}
Amos Fiat, Yuval Rabani, Yiftach Ravid, and Baruch Schieber.
\newblock A deterministic $o(k^3)$-competitive k-server algorithm for the
  circle.
\newblock {\em Algorithmica}, 11:572--578, 1994.

\bibitem{FiaRic94}
Amos Fiat and Moty Ricklin.
\newblock Competitive algorithms for the weighted server problem.
\newblock {\em Theoret. Comput. Sci.}, 130:85--99, 1994.

\bibitem{Grove91}
Edward Grove.
\newblock The harmonic $k$-server algorithm is competitive.
\newblock In {\em Proc. 23rd Symp. Theory of Computing (STOC)}, pages 260--266.
  ACM, 1991.

\bibitem{Herrma05}
Sven Herrmann.
\newblock Kombinatorik von {H}ypersimplex-{T}riangulierungen.
\newblock Master's thesis, Technische Universit{\"a}t Darmstadt, 2005.

\bibitem{HuKoMo04}
Katharina~T. Huber, Jacobus~H. Koolen, and Vincent Moulton.
\newblock The tight span of an antipodal metric space: Part {II} -- geometrical
  properties.
\newblock {\em Discrete {\&} Computational Geometry}, 31(4):567--586, 2004.

\bibitem{HuKoMo05}
Katharina~T. Huber, Jacobus~H. Koolen, and Vincent Moulton.
\newblock The tight span of an antipodal metric space: Part {I} --
  combinatorial properties.
\newblock {\em Discrete Mathematics}, 303(1-3):65--79, 2005.

\bibitem{HuKoMo06}
Katharina~T. Huber, Jacobus~H. Koolen, and Vincent Moulton.
\newblock On the structure of the tight-span of a totally split-decomposable
  metric.
\newblock {\em Eur. J. Comb.}, 27(3):461--479, 2006.

\bibitem{IraRub91}
Sandy Irani and R.~Rubinfeld.
\newblock A competitive 2-server algorithm.
\newblock {\em Inform. Process. Lett.}, 39:85--91, 1991.

\bibitem{Isbell64}
John~R. Isbell.
\newblock Six theorems about metric spaces.
\newblock {\em Comment. Math. Helv.}, 39:65--74, 1964.

\bibitem{KaRaRa94}
H.~Karloff, Yuval Rabani, and Yiftach Ravid.
\newblock Lower bounds for randomized k-server and motion-planning algorithms.
\newblock {\em SIAM J. Comput.}, 23:293--312, 1994.

\bibitem{KoMoTo98}
Jacobus Koolen, Vincent Moulton, and Udo T{\"o}nges.
\newblock The coherency index.
\newblock {\em Discrete Math.}, 192:205--222, 1998.

\bibitem{KoMoTo00}
Jacobus Koolen, Vincent Moulton, and Udo T{\"o}nges.
\newblock A classification of the six-point prime metrics.
\newblock {\em European J. Combinatorics}, 21:815--829, 2000.

\bibitem{Koutso94}
Elias Koutsoupias.
\newblock {\em On-line algorithms and the $k$-server conjuncture}.
\newblock PhD thesis, University of California, San Diego, CA, 1994.

\bibitem{Koutso99}
Elias Koutsoupias.
\newblock Weak adversaries for the $k$-server problem.
\newblock In {\em Proc. 40th Symp. Foundations of Computer Science (FOCS)},
  pages 444--449. IEEE, 1999.

\bibitem{KouPap94A}
Elias Koutsoupias and Christos Papadimitriou.
\newblock On the $k$-server conjecture.
\newblock In {\em Proc. 26th Symp. Theory of Computing (STOC)}, pages 507--511.
  ACM, 1994.

\bibitem{KouPap95A}
Elias Koutsoupias and Christos Papadimitriou.
\newblock On the $k$-server conjecture.
\newblock {\em J. ACM}, 42:971--983, 1995.

\bibitem{MaMcSl88}
Mark Manasse, Lyle~A. McGeoch, and Daniel Sleator.
\newblock Competitive algorithms for online problems.
\newblock In {\em Proc. 20th Symp. Theory of Computing (STOC)}, pages 322--333.
  ACM, 1988.

\bibitem{RagSni89}
Prabhakar Raghavan and Marc Snir.
\newblock Memory versus randomization in online algorithms.
\newblock In {\em Proc. 16th International Colloquium on Automata, Languages,
  and Programming (ICALP)}, volume 372 of {\em Lecture Notes in Comput. Sci.},
  pages 687--703. Springer, 1989.

\bibitem{RagSni94}
Prabhakar Raghavan and Marc Snir.
\newblock Memory versus randomization in on-line algorithms.
\newblock {\em IBM J. Res. Dev.}, 38:683--707, 1994.

\bibitem{SemSte03}
Charles Semple and Mike Steel.
\newblock {\em Phylogenetics}.
\newblock Oxford University Press, 2003.

\bibitem{SVTSDK96}
C.~Stock, B.~Volkmer, Udo T{\"o}nges, M.~Silva, Andreas W.~M. Dress, and
  Andreas Kr{\"a}mer.
\newblock Vergleichende {A}nalyse von {HTLV}-{I}-{N}ukleotidsequenzen mittels
  {S}plit-{Z}erlegungsmethode.
\newblock In {\em GMDS}, pages 533--537, 1996.

\bibitem{StuYu04}
Bernd Sturmfels and Josephine Yu.
\newblock Classification of six-point metrics.
\newblock {\em Electr. J. Comb.}, 11(1), 2004.
\newblock $http://www.combinatorics.org/Volume_11/Abstracts/v11i1r44.html$.

\bibitem{Teia93d}
Boris Teia.
\newblock Personal Communication.

\bibitem{Teia93b}
Boris Teia.
\newblock {\em Ein {B}eitrag zum k-{S}erver {P}roblem}.
\newblock PhD thesis, Universit{\"a}t des Saarlandes, 1993.

\bibitem{Young94}
Neal~E. Young.
\newblock The k-server dual and loose competitiveness for paging.
\newblock {\em Algorithmica}, 11:525--541, 1994.
\newblock Preliminary version appeared in SODA'91 titled ``On-Line Caching as
  Cache Size Varies''.

\end{thebibliography}

\eject
\begin{appendix}
\section{Appendix: Mathematica Calculations}
\label{app: mathematica}\label{app: math}

We used Mathematica 5.2 to rewrite the left hand side of Inequality
(\ref{harmonic update condition active}) as a single rational expression
with the least common denominator.  We then substituted zero for $efg$
throughout.  Then {\it numerator$_\ssb{1}$}, the numerator
of the resulting rational expression, is the following polynomial:
\vskip 0.05in
\noindent
{\footnotesize
$4 a^3 b c+9 a^2 b^2 c+5 a b^3 c+4 a^3 c^2+14 a^2 b c^2+
13 a b^2 c^2+3 b^3 c^2+5 a^2 c^3+9 a b c^3+4 b^2 c^3+a c^4+b c^4+
2 a^3 b d+3 a^2 b^2 d+a b^3 d+6 a^3 c d+23 a^2 b c d+24 a b^2 c d+
7 b^3 c d+17 a^2 c^2 d+36 a b c^2 d+17 b^2 c^2 d+10 a c^3 d+
9 b c^3 d+c^4 d+2 a^3 d^2+7 a^2 b d^2+5 a b^2 d^2+18 a^2 c d^2+
37 a b c d^2+17 b^2 c d^2+21 a c^2 d^2+20 b c^2 d^2+5 c^3 d^2+
6 a^2 d^3+10 a b d^3+4 b^2 d^3+16 a c d^3+16 b c d^3+8 c^2 d^3+
4 a d^4+4 b d^4+4 c d^4+4 a^3 c e+17 a^2 b c e+17 a b^2 c e+
4 b^3 c e+13 a^2 c^2 e+26 a b c^2 e+11 b^2 c^2 e+7 a c^3 e+6 b c^3 e+
c^4 e+2 a^3 d e+9 a^2 b d e+7 a b^2 d e+25 a^2 c d e+50 a b c d e+
21 b^2 c d e+30 a c^2 d e+26 b c^2 d e+7 c^3 d e+10 a^2 d^2 e+
18 a b d^2 e+6 b^2 d^2 e+37 a c d^2 e+34 b c d^2 e+16 c^2 d^2 e+
14 a d^3 e+14 b d^3 e+14 c d^3 e+4 d^4 e+8 a^2 c e^2+16 a b c e^2+
6 b^2 c e^2+10 a c^2 e^2+7 b c^2 e^2+2 c^3 e^2+6 a^2 d e^2+10 a b d e^2+
2 b^2 d e^2+22 a c d e^2+17 b c d e^2+9 c^2 d e^2+12 a d^2 e^2+
10 b d^2 e^2+13 c d^2 e^2+6 d^3 e^2+4 a c e^3+2 b c e^3+c^2 e^3+
4 a d e^3+2 b d e^3+3 c d e^3+2 d^2 e^3+2 a^3 b f+4 a^2 b^2 f+
2 a b^3 f+6 a^3 c f+30 a^2 b c f+33 a b^2 c f+8 b^3 c f+22 a^2 c^2 f+
46 a b c^2 f+21 b^2 c^2 f+13 a c^3 f+12 b c^3 f+c^4 f+4 a^3 d f+
16 a^2 b d f+13 a b^2 d f+2 b^3 d f+39 a^2 c d f+88 a b c d f+
43 b^2 c d f+54 a c^2 d f+54 b c^2 d f+13 c^3 d f+16 a^2 d^2 f+
29 a b d^2 f+10 b^2 d^2 f+59 a c d^2 f+59 b c d^2 f+31 c^2 d^2 f+
18 a d^3 f+16 b d^3 f+24 c d^3 f+4 d^4 f+2 a^3 e f+9 a^2 b e f+
8 a b^2 e f+b^3 e f+30 a^2 c e f+67 a b c e f+30 b^2 c e f+
41 a c^2 e f+39 b c^2 e f+10 c^3 e f+20 a^2 d e f+41 a b d e f+
14 b^2 d e f+90 a c d e f+85 b c d e f+45 c^2 d e f+42 a d^2 e f+
39 b d^2 e f+58 c d^2 e f+22 d^3 e f+5 a^2 e^2 f+9 a b e^2 f+
2 b^2 e^2 f+30 a c e^2 f+26 b c e^2 f+15 c^2 e^2 f+26 a d e^2 f+
21 b d e^2 f+37 c d e^2 f+23 d^2 e^2 f+3 a e^3 f+b e^3 f+6 c e^3 f+
7 d e^3 f+2 a^3 f^2+10 a^2 b f^2+10 a b^2 f^2+2 b^3 f^2+23 a^2 c f^2+
57 a b c f^2+28 b^2 c f^2+35 a c^2 f^2+35 b c^2 f^2+8 c^3 f^2+
16 a^2 d f^2+32 a b d f^2+12 b^2 d f^2+70 a c d f^2+74 b c d f^2+
40 c^2 d f^2+28 a d^2 f^2+24 b d^2 f^2+46 c d^2 f^2+12 d^3 f^2+
11 a^2 e f^2+25 a b e f^2+10 b^2 e f^2+57 a c e f^2+58 b c e f^2+
31 c^2 e f^2+44 a d e f^2+41 b d e f^2+74 c d e f^2+37 d^2 e f^2+
14 a e^2 f^2+12 b e^2 f^2+26 c e^2 f^2+26 d e^2 f^2+4 e^3 f^2+6 a^2 f^3+
14 a b f^3+6 b^2 f^3+29 a c f^3+32 b c f^3+17 c^2 f^3+20 a d f^3+
18 b d f^3+38 c d f^3+14 d^2 f^3+17 a e f^3+17 b e f^3+32 c e f^3+
27 d e f^3+10 e^2 f^3+6 a f^4+6 b f^4+12 c f^4+8 d f^4+8 e f^4+
2 f^5+4 a^3 b g+8 a^2 b^2 g+4 a b^3 g+8 a^3 c g+34 a^2 b c g+
33 a b^2 c g+6 b^3 c g+23 a^2 c^2 g+45 a b c^2 g+19 b^2 c^2 g+
13 a c^3 g+12 b c^3 g+c^4 g+6 a^3 d g+24 a^2 b d g+23 a b^2 d g+
6 b^3 d g+47 a^2 c d g+102 a b c d g+49 b^2 c d g+58 a c^2 d g+
57 b c^2 d g+13 c^3 d g+23 a^2 d^2 g+44 a b d^2 g+18 b^2 d^2 g+
67 a c d^2 g+66 b c d^2 g+33 c^2 d^2 g+22 a d^3 g+20 b d^3 g+
26 c d^3 g+4 d^4 g+4 a^3 e g+17 a^2 b e g+16 a b^2 e g+3 b^3 e g+
34 a^2 c e g+71 a b c e g+30 b^2 c e g+42 a c^2 e g+39 b c^2 e g+
10 c^3 e g+28 a^2 d e g+57 a b d e g+22 b^2 d e g+100 a c d e g+
93 b c d e g+46 c^2 d e g+51 a d^2 e g+47 b d^2 e g+61 c d^2 e g+
24 d^3 e g+9 a^2 e^2 g+17 a b e^2 g+6 b^2 e^2 g+32 a c e^2 g+
27 b c e^2 g+14 c^2 e^2 g+30 a d e^2 g+24 b d e^2 g+37 c d e^2 g+
24 d^2 e^2 g+5 a e^3 g+3 b e^3 g+5 c e^3 g+6 d e^3 g+6 a^3 f g+
31 a^2 b f g+32 a b^2 f g+7 b^3 f g+55 a^2 c f g+125 a b c f g+
56 b^2 c f g+72 a c^2 f g+69 b c^2 f g+16 c^3 f g+47 a^2 d f g+
99 a b d f g+44 b^2 d f g+162 a c d f g+167 b c d f g+84 c^2 d f g+
76 a d^2 f g+69 b d^2 f g+102 c d^2 f g+28 d^3 f g+26 a^2 f^2 g+
62 a b f^2 g+28 b^2 f^2 g+99 a c f^2 g+103 b c f^2 g+52 c^2 f^2 g+
86 a d f^2 g+83 b d f^2 g+128 c d f^2 g+55 d^2 f^2 g+34 a f^3 g+
35 b f^3 g+53 c f^3 g+45 d f^3 g+14 f^4 g+4 a^3 g^2+19 a^2 b g^2+
18 a b^2 g^2+3 b^3 g^2+31 a^2 c g^2+65 a b c g^2+26 b^2 c g^2+
37 a c^2 g^2+34 b c^2 g^2+8 c^3 g^2+30 a^2 d g^2+64 a b d g^2+
30 b^2 d g^2+92 a c d g^2+93 b c d g^2+45 c^2 d g^2+48 a d^2 g^2+
45 b d^2 g^2+57 c d^2 g^2+16 d^3 g^2+20 a^2 e g^2+43 a b e g^2+
17 b^2 e g^2+67 a c e g^2+63 b c e g^2+33 c^2 e g^2+70 a d e g^2+
66 b d e g^2+87 c d e g^2+49 d^2 e g^2+22 a e^2 g^2+19 b e^2 g^2+
28 c e^2 g^2+32 d e^2 g^2+5 e^3 g^2+33 a^2 f g^2+76 a b f g^2+
33 b^2 f g^2+109 a c f g^2+107 b c f g^2+53 c^2 f g^2+110 a d f g^2+
109 b d f g^2+143 c d f g^2+68 d^2 f g^2+64 a f^2 g^2+65 b f^2 g^2+
86 c f^2 g^2+86 d f^2 g^2+35 f^3 g^2+13 a^2 g^3+28 a b g^3+
11 b^2 g^3+39 a c g^3+36 b c g^3+18 c^2 g^3+44 a d g^3+44 b d g^3+
53 c d g^3+27 d^2 g^3+31 a e g^3+29 b e g^3+39 c e g^3+46 d e g^3+
15 e^2 g^3+50 a f g^3+49 b f g^3+61 c f g^3+69 d f g^3+41 f^2 g^3+
14 a g^4+13 b g^4+16 c g^4+20 d g^4+15 e g^4+23 f g^4+5 g^5$\/}
\vskip 0.1in

\noindent
Each of the variables $a,b,c,d,e,f,g$ is an isolation index, hence
cannot be negative.  Since there are no negative coefficients
in the polynomial $numerator_\ssb{1}$, its value must be non-negative.
\noindent
\vskip 0.05in

\eject
Similarly, we used Mathematica to find a polynomial expression for
$numerator_\ssb{2}$, the numerator of the left hand
side of Inequality (\ref{harmonic update condition cryptic}):
\vskip 0.05in
\noindent
{\footnotesize
$4 a^2 b+4 a b^2+9 a b c+4 b^2 c+6 b c^2+2 a^2 d+6 a b d+4 b^2 d+
3 a c d+6 b c d+2 a d^2+2 b d^2+2 c d^2+2 a^2 e+11 a b e+4 b^2 e+
2 a c e+12 b c e+6 a d e+9 b d e+3 c d e+2 d^2 e+3 a e^2+7 b e^2+
c e^2+3 d e^2+e^3+8 a b f+4 b^2 f+8 b c f+4 a d f+6 b d f+
4 c d f+2 d^2 f+5 a e f+11 b e f+3 c e f+7 d e f+4 e^2 f+4 b f^2+
2 d f^2+3 e f^2+4 a^2 g+12 a b g+4 b^2 g+10 a c g+15 b c g+
6 c^2 g+7 a d g+9 b d g+6 c d g+2 d^2 g+12 a e g+15 b e g+
12 c e g+9 d e g+7 e^2 g+7 a f g+11 b f g+9 c f g+7 d f g+
3 f^2 g+9 a g^2+9 b g^2+11 c g^2+5 d g^2+11 e g^2+8 f g^2+5 g^3$\/}
\vskip 0.1in
\noindent
Every term is non-negative, hence $numerator_\ssb{2}$ is
non-negative.
\end{appendix}

\end{document}